\documentclass[pdflatex,sn-mathphys-ay,iicol]{sn-jnl}
\usepackage[abs]{overpic}
\usepackage{tabularx}
\usepackage{graphicx}%
\usepackage{multirow}%
\usepackage{amsmath,amssymb,amsfonts}%
\usepackage{amsthm}%
\usepackage{mathrsfs}%
\usepackage[title]{appendix}%
\usepackage{xcolor}%
\usepackage{textcomp}%
\usepackage{manyfoot}%
\usepackage{booktabs}%
\usepackage{algorithm}%
\usepackage{algorithmicx}%
\usepackage{algpseudocode}%
\usepackage{listings}%
\usepackage{scalerel}
\usepackage{tikz}
\usetikzlibrary{svg.path}
\usetikzlibrary{calc} 
\usepackage[utf8]{inputenc}
\usepackage[english]{babel}
\usepackage{hyperref}
\usepackage{natbib}
\usepackage{url}
\setcitestyle{authoryear,round}
\usepackage{svg}
\usepackage{upgreek}
\usepackage[dvipsnames]{xcolor}
\usepackage{subcaption}
\usepackage{float}
\usepackage{comment}
\usepackage[normalem]{ulem}
\usetikzlibrary{arrows.meta,positioning,fit,backgrounds,calc}

\newcommand\thefont{\expandafter\string\the\font}

\newcommand\orcidicon[1]{\href{https://orcid.org/my-orcid?orcid=0009-0007-0885-6689}
{\mbox{\scalerel*{
\begin{tikzpicture}[yscale=-1,transform shape]
\pic{orcidlogo};
\end{tikzpicture}
}{|}}}}
\newcommand{\commentout}[1]{}

\definecolor{darkgreen}{rgb}{0.0, 0.5, 0.0}
\definecolor{orcidlogocol}{HTML}{A6CE39}

\tikzset{
  orcidlogo/.pic={
    \fill[orcidlogocol] svg{M256,128c0,70.7-57.3,128-128,128C57.3,256,0,198.7,0,128C0,57.3,57.3,0,128,0C198.7,0,256,57.3,256,128z};
    \fill[white] svg{M86.3,186.2H70.9V79.1h15.4v48.4V186.2z}
                 svg{M108.9,79.1h41.6c39.6,0,57,28.3,57,53.6c0,27.5-21.5,53.6-56.8,53.6h-41.8V79.1z M124.3,172.4h24.5c34.9,0,42.9-26.5,42.9-39.7c0-21.5-13.7-39.7-43.7-39.7h-23.7V172.4z}
                 svg{M88.7,56.8c0,5.5-4.5,10.1-10.1,10.1c-5.6,0-10.1-4.6-10.1-10.1c0-5.6,4.5-10.1,10.1-10.1C84.2,46.7,88.7,51.3,88.7,56.8z};
  }
}

\usepackage{acronym}
    \acrodef{AFC}[AFC]{Active Flow Control}
    \acrodef{BO}[BO]{Bayesian Optimization}
    \acrodef{BOS}[BOS]{Background Oriented Schlieren Method}
    \acrodef{DEHS}[DEHS]{Di-Ethyl-Hexyl-Sebacate}
    \acrodef{DR}[DR]{Dynamic Range}
    \acrodef{EBIV}[EBIV]{Event-Based Imaging Velocimetry}
    \acrodef{EBV}[EBV]{Event-Based Vision}
    \acrodef{EI}[EI]{Expected Improvement}
    \acrodef{FPGA}[FPGA]{Field Programmable Gate Array}
    \acrodef{GP}[GP]{Gaussian Process}
    \acrodef{GPU}[GPU]{Graphics Processing Units}
    \acrodef{HR}[HR]{High-Resolution}
    \acrodef{IW}[IW]{Interrogation Window}
    \acrodef{ISOMAP}[ISOMAP]{Isometric mapping}
    \acrodef{KF}[KF]{Kalman Filter}
    \acrodef{LOR}[LOR]{Low-Order Reconstruction}
    \acrodef{LR}[LR]{Low-Resolution}
    \acrodef{LSE}[LSE]{Linear Stochastic Estimation}
    \acrodef{PIV}[PIV]{Particle Image Velocimetry}
    \acrodef{POD}[POD]{Proper Orthogonal Decomposition}
    \acrodef{PTV}[PTV]{Particle Tracking Velocimetry}
    \acrodef{PSD}[PSD]{Power Spectral Density}
    \acrodef{RBF}[RBF]{Radial Basis Function}
    \acrodef{RMSE}[RMSE]{Root-Mean Squared Error}
    \acrodef{ROI}[ROI]{Region of Interest}
    \acrodef{ROM}[ROM]{Reduced-Order Modeling}
    \acrodef{rtEBIV}[rt-EBIV]{real-time Event-Based Imaging Velocimetry}
    \acrodef{SPOD}[SPOD]{Spectral Proper Orthogonal Decomposition}
    \acrodef{TKE}[TKE]{Turbulent Kinetic Energy}
    \acrodef{VR}[VR]{Variance Rescaling}
\usepackage{placeins}
\definecolor{RevOneCol}{rgb}{0.0,0.0,0.85}
\definecolor{RevTwoCol}{rgb}{0.85,0.0,0.0}
\definecolor{RevThreeCol}{rgb}{0.0,0.55,0.0}
\newcommand{\revA}[1]{\textcolor{RevOneCol}{#1}}
\newcommand{\revB}[1]{\textcolor{RevTwoCol}{#1}}
\newcommand{\revC}[1]{\textcolor{RevThreeCol}{#1}}
\newcommand{\checkme}[1]{\textcolor{orange}{\textbf{[CHECK: }#1\textbf{]}}}
\usepackage[mathlines]{lineno}
\newif\ifcleanversion
\cleanversiontrue   % <-- change to \cleanversiontrue for the clean copy
\ifcleanversion
  \renewcommand{\revA}[1]{#1}
  \renewcommand{\revB}[1]{#1}
  \renewcommand{\revC}[1]{#1}
  \renewcommand{\checkme}[1]{}
  \renewcommand{\sout}[1]{}
  \renewcommand{\linenumbers}{}
\fi
\begin{document}

\title[Article Title]{Real-Time Estimation of High-Resolution Flow Fields and Reduced-Order Coordinates from Event-Based Imaging Velocimetry}

%%===========================================================================================%%
\author*[1]{\fnm{Luca} \sur{Franceschelli}}\email{luca.franceschelli@uc3m.es \orcidicon{}}

\author[2]{\fnm{Enrico} \sur{Amico }}
\author[3]{\fnm{Christian E.} \sur{Willert}}
\author[1]{\fnm{Marco} \sur{Raiola}}
\author[2]{\fnm{Gioacchino} \sur{Cafiero}}
\author[1]{\fnm{Stefano} \sur{Discetti}}


\affil*[1]{\orgdiv{Department of Aerospace Engineering}, \orgname{Universidad Carlos III de Madrid}, \orgaddress{\street{Avda. Universidad 30}, \city{Leganés}, \postcode{28911}, \state{Madrid}, \country{Spain}}}

% \affil[2]{\orgdiv{}, \orgname{}, \city{}, \country{}}
% \affil[3]{\orgdiv{}, \orgname{}, \city{}, \country{}}

\affil*[2]{\orgdiv{Department of Mechanical and Aerospace Engineering}, \orgname{Politecnico di Torino}, \orgaddress{\street{Corso Duca degli Abruzzi 24}, \city{Turin}, \postcode{10129}, \country{Italy}}}

\affil*[3]{\orgdiv{DLR Institute of Propulsion Technology}, \orgname{German Aerospace Center}, \orgaddress{\street{Linder Höhe}, \city{Köln}, \postcode{51170},  \country{Germany}}}

%%===========================================================================================%%
\abstract{\acresetall
A data-driven framework is proposed to estimate \ac{HR} velocity fields and reduced-order flow coordinates from \ac{rtEBIV} measurements. \ac{LR} data are retrieved through fast event analysis on a coarse grid. During offline training, the \ac{LR}/\ac{HR} mapping and a linear dynamical model in a \ac{POD}-based reduced-order space are identified. Online, each incoming \ac{LR} snapshot is projected onto the \ac{LR} basis, and the corresponding \ac{HR} reduced-order coordinates are estimated and temporally regularized. Three estimators are proposed and compared: a direct \ac{KF}, a \ac{LSE} combined with Kalman filtering, and a variance-rescaled implementation (LSE+VR). The full \ac{HR} velocity field is then reconstructed from the estimated reduced-order coordinates using the retained \ac{POD} basis.
The framework is assessed on two turbulent flows acquired by pulsed \ac{EBIV}: a submerged water jet and a channel flow over a square rib. All three estimators outperform direct cubic interpolation of the \ac{LR} fields, recovering physically consistent \ac{HR} flow states with improved estimation of turbulent kinetic energy, spectral content, reduced-order dynamics, and temporal coherence. The linear stochastic estimator yields the lowest overall reconstruction error, while its variance-rescaled implementation provides nearly identical values and improves the recovery of fluctuation energy and higher-order content. The direct \ac{KF} formulation remains highly efficient in terms of computational cost and provides the closest agreement with the \ac{HR} reference in the temporal and spatial spectral analyses.
The bulk of the computational cost lies in the full-field \ac{HR} reconstruction, whereas the reduced-order coordinate estimation alone is negligible compared with the cost of the \ac{LR} processing itself. The framework enables deliberately coarse real-time \ac{EBIV} processing to be combined with reduced-order refinement, extending the effective real-time operating range toward higher update frequencies while still providing richer and dynamically more consistent \ac{HR} flow representations. This makes the approach attractive for high-quality real-time flow diagnostics and as a first step toward future observer-based and closed-loop flow-control applications.}


\maketitle
\linenumbers
\acresetall

\clearpage

%%===========================================================================================%%

\section{Introduction}\label{sec1}
\acf{PIV} is widely used for experimental flow diagnostics because it provides non-intrusive, spatially resolved velocity measurements over an extended field of view \citep{raffelParticleImageVelocimetry2018}. Compared with pointwise sensing, full-field velocimetry provides direct access to the spatial distribution of the velocity field and to quantities that depend on it, such as gradients, shear, vorticity, and multi-point correlations. This information is central to physical interpretation and particularly valuable for reduced-order modelling and observer design, since it enables state descriptions that are linked to the spatial structure of the flow rather than inferred from a small number of local probes \citep{glauser1987,rajaee1994,berkooz2003}. For flow control applications, spatially distributed velocity information is particularly attractive because it supports state representations that are more directly connected to the underlying dynamics than those inferred from a limited number of local measurements \citep{siegelRealTimeParticleImage2003, willertRealTimeParticleImage2010,gautierClosedloopSeparationControl2015,varonAdaptiveControlDynamics2019, mccormickReactiveControlVelocity2024, nonomura2025}.

% However, despite \ac{PIV} provides detailed spatially resolved information, it comes with a limited temporal bandwidth and a higher computational latency which is typically not compatible with real-time applications. On the other hand, conventional pointwise probes --- such as hot wires, pressure transducers, or microphones --- offer high sampling rates, straightforward hardware integration, and minimal processing requirements but only provide sparse spatial information and often at the cost of high intrusiveness. Flow control experiments, where real-time information is critical, have traditionally relied on point sensors (see e.g., \citealp{audiffredReactiveExperimentalControl2024,dacome_opposition_2024}), although their effective use generally requires careful sensor placement informed by prior knowledge of the flow and, in practice, substantial experimental expertise. Real-time imaging velocimetry could bridge this gap by combining the spatial information of velocity field measurements with the responsiveness needed for monitoring, estimation, and, ultimately, control.

While \ac{PIV} provides detailed spatially resolved information, it comes with limited temporal bandwidth and higher computational latency. In contrast, conventional probes (such as hot wires, pressure transducers, or microphones) are more suitable for real-time applications due to their high sampling rates, straightforward hardware integration, and minimal processing requirements. However, they provide only sparse spatial information and are often intrusive. Flow-control experiments have traditionally relied on point sensors (see e.g., \citealp{audiffredReactiveExperimentalControl2024,dacomeOppositionFlowControl2024}), although their effective use generally requires careful sensor placement informed by prior knowledge of the flow, even in canonical configurations. Real-time imaging velocimetry could bridge this gap by combining the spatial information of velocity-field measurements with the responsiveness needed for monitoring, estimation, and, ultimately, control. \revA{\footnote{\revA{Throughout this
work, the terms \emph{online} and \emph{real-time} are used interchangeably to denote processing performed concurrently with the experiment, at low latency. We note that, strictly speaking, real-time operation implies meeting a hard deadline dictated by the application \citep{stankovic2002misconceptions}, whereas online operation only implies immediate, non-batch processing; the distinction is application-dependent and the relevant latency requirement
here is set by the flow time scales to be resolved.}}}

The attractiveness of implementing field information in flow control strategies has motivated several studies aimed at advancing \ac{PIV} toward real-time operation, including accelerated image-processing pipelines \citep{kreizerRealtimeImageProcessing2010, willertRealTimeParticleImage2010}, optical-flow-based velocimetry \citep{gautierClosedloopSeparationControl2015,pimientaHighResolutionHighSpeed2025, bolltRapidPIVFullFlowField2025}, and sparse or reduced-order reconstruction strategies \citep{kandaProofofconceptStudySparse2022}. However, real-time operation is generally obtained at the cost of either specialized hardware and tailored implementations, such as \ac{FPGA}-based processing \citep{kreizerRealtimeImageProcessing2010} or \ac{GPU} acceleration \citep{pimientaHighResolutionHighSpeed2025}, or a simplified measurement and processing chain. In the latter case, latency is reduced by accepting compromises such as coarser spatial resolution, fewer iterations, or lighter validation.  \cite{kandaProofofconceptStudySparse2022} proposed sparse processing PIV (SPPIV) as a framework for real-time flow field estimation. Rather than processing the full image, cross-correlation is performed only at a small number of interrogation windows placed at optimally selected spatial locations. A pre-trained \ac{KF}, formulated in the \ac{POD} coefficient space, maps the resulting sparse velocity vectors onto a low-dimensional flow representation, from which the full velocity field is reconstructed. The method estimated a small number of modes owing to the limited number of simultaneous sparse measurements. The effectiveness of this approach for closed-loop flow control has been subsequently demonstrated by \cite{nonomura2025} and \citet{viguera2026}\revA{, who employed the SPPIV-estimated velocity field as a real-time feedback signal at update frequencies of up to 2 kHz for plasma-actuator-based control. Their studies respectively achieved regulation of the velocity field toward prescribed targets and mitigation of flow separation over an airfoil}.

\ac{EBV} provides a new route to perform cost-effective flow field measurements in the \revB{kHz} range. \ac{EBV} replaces conventional frame-based acquisition with asynchronous brightness-change detection at the pixel level \citep{mahowaldSiliconRetina1991,lichtsteiner128times128120DB2008}. An event sensor outputs a stream of sparse spatiotemporal events only when the local intensity change crosses a contrast threshold. \revA{Each event carries a polarity flag: positive events are triggered by a local brightness increase exceeding the contrast threshold, whereas negative events correspond to a brightness decrease.} A comprehensive review of this sensing paradigm and applications is provided by \cite{gallegoEventBasedVisionSurvey2022}. Compared with frame-based cameras, event sensors offer major advantages in terms of temporal responsiveness, low redundancy, reduced data transfer, and high dynamic range, making them attractive for flow diagnostics with real-time capability.

The first application of an \ac{EBV} sensor to velocimetry was reported by \cite{drazenRealtimeParticleTracking2011}, who demonstrated particle tracking in a 5-cm-diameter perspex pipe. Over the following decade, event cameras found use across a broad range of flow measurement scenarios, including \ac{BOS} \citep{shibaEventBasedBackgroundOrientedSchlieren2024}, microfluidics \citep{howell2020}, and cardiac flows \citep{amico2026}. \cite{RuschPhdThesis2023} and \cite{ruschTrackAERRealtimeEventbased2023} developed a real-time system for three-dimensional particle tracking at relatively high velocities, albeit with a hard upper bound on the number of tracks that can be reliably reconstructed. More recently, \cite{willertEventbasedParticleImage2025} proposed a dual-sensor approach that yields dense and accurate velocity fields even in high-speed flows\revA{, with demonstrated measurements in a free jet with exit velocities up to 180 m/s at laser pulse separations down to 100~ns}.

%At the same time, \ac{EBV} introduces specific limitations. The scene is described in a binary fashion, retaining only the sign of local intensity changes. Moreover, in dense particle fields or under sustained illumination, the asynchronous event traffic can become intense enough to saturate the internal arbiter — the on-chip component responsible for assigning a timestamp to each pixel activation — thereby degrading measurement quality.

%Pulsed-\ac{EBIV}, referred to hereafter simply as \ac{EBIV}, addresses temporal jitter and arbiter saturation by combining event-based sensing with pulsed illumination \citep{willertEventbasedImagingVelocimetry2023}. Each laser pulse generates a compact burst of particle-induced events confined to a short temporal window, limiting the instantaneous event load on the sensor. The resulting event stream can then be partitioned into sparse binary pseudo-images suitable for correlation-based velocimetry, while retaining the low-latency and low-redundancy advantages of \ac{EBV}. 

\cite{franceschelliAssessmentEventbasedImaging2025} have shown that \ac{EBIV} measurements can support reduced-order flow estimation in turbulent configurations. \cite{willert2026}  demonstrated the use of \ac{EBIV} for real-time jet-flow control,  reaching real-time full-frame measurements at an average rate of $270$~Hz with a single-pass cross-correlation analysis on pseudo-frames, thus partially sacrificing spatial resolution. This opens the following question: \revA{\textbf{can coarse, data-efficient \ac{EBIV} measurements be used to estimate, in real time, high-resolution velocity fields and their reduced-order coordinates, without requiring high-speed frame-based cameras, high-power laser illumination, or the computational cost of a full offline-grade processing pipeline?}}

The present work addresses this question through a reduced-order cross-resolution estimation framework. A key point is that the \ac{LR} and \ac{HR} fields do not come from different instruments or different acquisition systems. Both datasets are extracted from the same raw \ac{EBV} event stream processed according to two different strategies: a lightweight single-pass procedure representative of real-time operation provides the \ac{LR} input, whereas a more accurate offline multi-frame, multi-pass procedure provides the \ac{HR} reference used for training and assessment. No conventional frame-based camera data is used at any stage.

The proposed real-time estimation methodology combines this paired \ac{LR}/\ac{HR} processing with \ac{POD}-based reduced-order modelling. In the online estimation phase, each incoming \ac{LR} snapshot is projected onto the full-rank \ac{LR} basis and used to estimate the corresponding reduced-order \ac{HR} state. Three estimators are considered to map the \ac{LR} state into the \ac{HR} state: a direct \acf{KF} formulation, a \ac{LSE}-based estimator followed by Kalman filtering, and a variance-rescaled \ac{LSE} variant (\ac{LSE}+\ac{VR}) designed to mitigate the attenuation of poorly observable modes. The reconstructed fields are compared against the reduced-order \ac{HR} reference and against direct cubic interpolation of the \ac{LR} field onto the \ac{HR} grid, adopted as a purely spatial baseline.

In this sense, the \ac{LR} measurements can be regarded as spatially distributed, non-intrusive probes that can be combined with data-driven methods to reconstruct the corresponding \ac{HR} flow field \citep{tu2013,discetti2019}.

\revA{The framework is evaluated on two broadband turbulent flows: a submerged water jet at $Re_D = U_j D/\nu \approx 6,000$, and a channel flow over a square rib at $Re_H = U_{ch} H/\nu \approx 1,500$. Here, $U_j$ is the jet bulk velocity, $U_{ch}$ is the asymptotic channel-flow velocity, $D$ is the nozzle-exit diameter, $H$ is the rib height, and $\nu$ is the kinematic viscosity}. These two cases were selected because they exhibit energetically distributed, multi-scale dynamics and therefore provide a demanding testbed for cross-resolution estimation. In addition to instantaneous fields, the analysis includes mean and second-order statistics, temporal and spatial spectra, reduced-order modal dynamics, temporal autocorrelation, overall reconstruction error, and computational cost.

The complete measurement and estimation workflow presented in this work is released as \emph{VIBEstream}, an open-source Python package providing a simple interface for event-camera recording, live streaming, quick \ac{PIV}-style processing, and real-time \ac{EBIV} with direct velocity-field output (see the code availability statement).

The remainder of the paper is organized as follows. Section~\ref{sec:methodology} introduces the common reduced-order formulation and the three estimation methods. Section~\ref{sec:exp_setup} describes the experimental datasets and the \ac{EBIV} processing chain used to generate the paired \ac{LR} and \ac{HR} fields. Section~\ref{sec:results} presents the reconstruction results and the computational-cost analysis. Finally, Section~\ref{sec:Conclusions} summarizes the work and discusses the practical implications, limitations, and possible extensions of the proposed framework.

% ============================================================
%  SECTION 2 — Methodology
% ============================================================
\section{Methodology}
\label{sec:methodology}
\begin{figure*}
    \centering
    \includegraphics{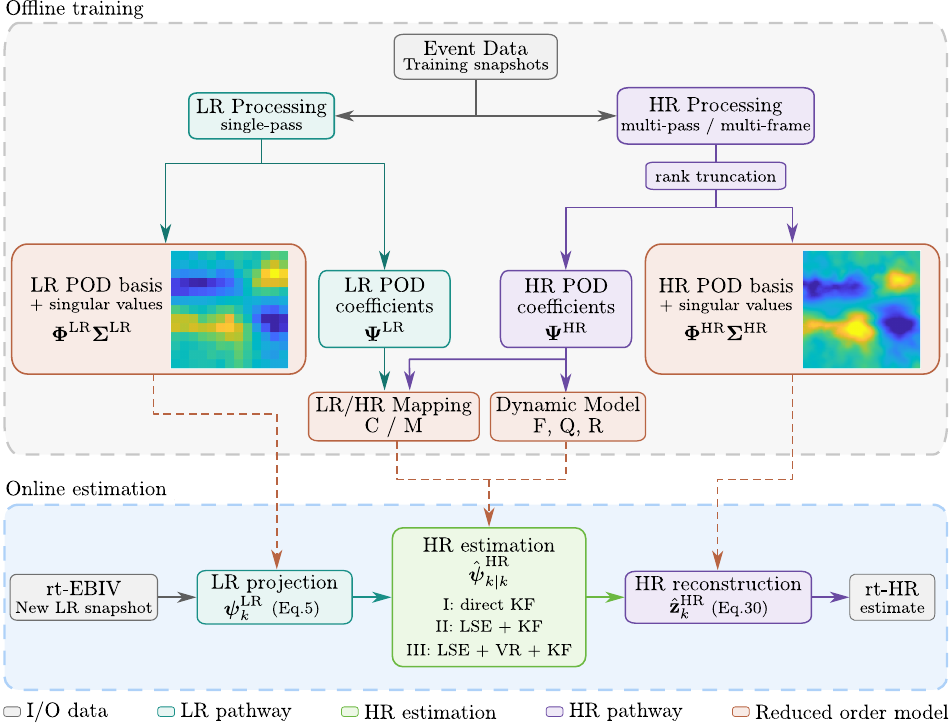}
    \caption{\revA{Offline training (top, gray dashed frame) and online estimation (bottom, blue dashed frame) pipeline. The color coding distinguishes input/output data (gray), the \ac{LR} pathway (teal), the \ac{HR} pathway (purple), the reduced-order model operators (orange), and the \ac{HR} estimation stage (green); dashed orange arrows denote the operators identified offline and deployed online. The abbreviation ``rt'' stands for real-time.}}
    \label{fig:pipeline}
\end{figure*}

This section describes the three estimation methodologies considered in the present work. Although they differ in the way the \ac{HR} coefficients are estimated and temporally regularized, all three methods share the same offline/online framework and use the temporal coefficients from \ac{LR} measurements as input.

\revA{For clarity, the main terminology used throughout the paper is defined here. The \emph{latent coordinates} are the \ac{POD} temporal coefficients, i.e. the modal amplitudes. The \emph{reduced-order coordinates} denote the subset of these coefficients retained after truncation of the \ac{POD} basis. Accordingly, the \emph{latent space}, or equivalently the \emph{reduced-order space}, is the subspace spanned by the retained \ac{POD} modes. In this work, \emph{state estimation} refers to the estimation of the retained \ac{POD} coefficients from coarse real-time measurements. Finally, the \emph{modal energy} denotes the \ac{POD} modal energy, defined as the squared singular value associated with each mode.}

%\revA{Throughout this paper, \emph{real-time} denotes processing for which the per-snapshot latency remains below the acquisition interval, allowing the measurement chain to sustain the acquisition rate without accumulating delay. By contrast, \emph{online} denotes operations performed sequentially on each incoming snapshot during the experiment, rather than through batch processing after acquisition. Online operation is therefore a prerequisite for real-time operation, whereas real-time operation additionally imposes a latency constraint.}

\revA{The workflow pipeline is reported in Figure \ref{fig:pipeline}.} The main structure is divided into two phases:
\begin{itemize}
    \item \textbf{Offline training}: event sequences are analyzed both with a simple coarse analysis based on single-pass cross-correlation with non-overlapping sampling window (\ac{LR} dataset) and with an advanced multi-pass multi-frame process (\ac{HR} dataset). A mapping is then established between the \ac{LR} and \ac{HR} data with the aid of reduced-order coordinates.
    \item \textbf{Online estimation} \revA{\sout{(real-time)}}: the stream of events is processed with the same quick evaluation process used to obtain the \ac{LR} dataset. The data is then used to estimate the \ac{HR} velocity fields using the mapping learned during training.
\end{itemize}

\revA{More explicitly, in the offline phase (top, gray dashed frame of Figure~\ref{fig:pipeline}) the recorded event data are processed by both pipelines: the single-pass \ac{LR} processing and the multi-pass, multi-frame \ac{HR} processing, the latter followed by the rank truncation discussed in Section~\ref{sec:formulation}. \ac{POD} applied to the two datasets yields the \ac{LR} basis and singular values $\boldsymbol{\Phi}^{\mathrm{LR}}\boldsymbol{\Sigma}^{\mathrm{LR}}$ with the associated coefficients $\boldsymbol{\Psi}^{\mathrm{LR}}$, and their truncated \ac{HR} counterparts. From the paired coefficient histories, the \ac{LR}/\ac{HR} mapping operator and the dynamics-model matrices are identified. In the online phase (bottom, blue dashed frame), each new \ac{LR} snapshot delivered by the real-time \ac{EBIV} processing is projected onto the \ac{LR} basis, Eq.~\eqref{eq:lr_projection}; the reduced-order \ac{HR} state $\hat{\boldsymbol{\psi}}^{\mathrm{HR}}_{k|k}$ is then estimated through one of the three methods of the following sections, using the operators learned offline (dashed orange arrows in the figure); finally, when the full field is required, the \ac{HR} snapshot is reconstructed through the retained \ac{HR} basis.}

\revA{From a practical standpoint, once the training is completed, the online stage operates as a self-contained estimator: each incoming \ac{LR} snapshot is converted, within the available time budget, into an estimate of the reduced-order \ac{HR} coordinates and, if required, into the corresponding full \ac{HR} field. In a prospective observer-based or closed-loop flow-control architecture, the estimated reduced-order coordinates would play the role of the plant state: they could be fed directly to a state-feedback controller or serve as the measurement input of a higher-level observer, without requiring the full-field reconstruction in the loop.}

A key point of the present framework is that the \ac{LR} and \ac{HR} velocity fields do not originate from different measurement systems. In both datasets, the same raw \ac{EBV} event stream is processed through two distinct velocimetry pipelines of different computational cost and accuracy. The first is a lightweight single-pass procedure designed to emulate a realistic real-time implementation and provides the \ac{LR} input to the estimator. The second is a more accurate offline multi-frame, multi-pass procedure and provides the \ac{HR} reference fields used for training, validation, and performance assessment. No conventional frame-based camera data is used at any stage of the present work.

The main hypothesis is that a \ac{LR}-\ac{HR} mapping can be established by using a compact mapping in latent coordinates. This approach is in line with previous works exploiting reduced-order representations and/or nonlinear mappings for super-resolution \citep{deng2019,fukami2021,cortina-fernandez2021,tirelli2023}.

In this section, the common formulation is introduced first, followed by the specific description of each method. First, the \ac{POD} implementation is briefly summarized, as it is used to identify the temporal coordinates for the mapping between \ac{LR} and \ac{HR} measurements, and for the state transitions. Later, the implementation of state observers for \ac{HR} full-state estimation is presented.

\subsection{Proper Orthogonal Decomposition and shared framework}
\label{sec:formulation}

Following the snapshot method introduced by \citet{sirovich1987}, \ac{POD} is applied to the \ac{HR} training dataset in fluctuation form. We denote the instantaneous velocity field as
\begin{equation}
\label{eq:vel_decomp}
\begin{aligned}
&\mathbf{U}(\mathbf{x},t)=
\overline{\mathbf{U}}(\mathbf{x})+\mathbf{u}(\mathbf{x},t)\,,
\end{aligned}
\end{equation}

\noindent
where $\overline{\mathbf{U}}(\mathbf{x})=\langle \mathbf{U}(\mathbf{x},t)\rangle_T$ denotes the temporal mean field and $\mathbf{u}(\mathbf{x},t)$ the corresponding fluctuation field according to the Reynolds decomposition. 
In PIV measurements, each snapshot samples the $c$ components of the fluctuating velocity field over a grid of sampling points covering the measurement domain. Let $N_p^{\mathrm{HR}}$ and $N_p^{\mathrm{LR}}$ denote the number of spatial sampling points of the \ac{HR} and \ac{LR} velocity fields, respectively. Then, $\mathbf{z}^{\mathrm{HR}}_k \in \mathbb{R}^{c\,N_p^{\mathrm{HR}}}$ and $\mathbf{z}^{\mathrm{LR}}_k \in \mathbb{R}^{c\,N_p^{\mathrm{LR}}}$ denote the corresponding $k$-th \ac{HR} and \ac{LR} fluctuation snapshots. Each snapshot is formed by stacking all the velocity components over all the available sampling points into a single column vector.
Denoting by $\mathbf{Z}^{\mathrm{HR}} \in \mathbb{R}^{c N_p^{\mathrm{HR}} \times N_t}$ the matrix collecting the $N_t$ \ac{HR} training snapshots, its economy-sized singular value decomposition is
\begin{equation}
\mathbf{Z}^{\mathrm{HR}}
=
\left[
\mathbf{z}^{\mathrm{HR}}_1,\ldots,\mathbf{z}^{\mathrm{HR}}_{N_t}
\right]
=
\mathbf{\Phi}^{\mathrm{HR}}
\mathbf{\Sigma}^{\mathrm{HR}}
\mathbf{\Psi}^{\mathrm{HR}^{\scriptstyle \mathsf{T}}},
\end{equation}
where $\mathbf{\Phi}^{\mathrm{HR}} \in \mathbb{R}^{c N_p^{\mathrm{HR}} \times m}$ contains the spatial modes, $\mathbf{\Psi}^{\mathrm{HR}} \in \mathbb{R}^{N_t \times m}$ the temporal modes, $\mathbf{\Sigma}^{\mathrm{HR}} \in \mathbb{R}^{m \times m}$ the singular values, and $m \le \min(N_t, c N_p^{\mathrm{HR}})$ the maximum rank of $\mathbf{Z}^{\mathrm{HR}}$.
A reduced-order representation of the \ac{HR} field can be obtained through a \ac{LOR} by truncating the \ac{HR} latent space from rank $m$ to rank $r$, retaining only the first $r$ modes:
\begin{equation}
    \mathbf{Z}^{\mathrm{HR}} \approx
    \boldsymbol{\Phi}^{\mathrm{HR}}_r
    \boldsymbol{\Sigma}^{\mathrm{HR}}_r
    \boldsymbol{\Psi}^{\mathrm{HR}^{\scriptstyle \mathsf{T}}}_r\,,
    \label{eq:hr_lor}
\end{equation}
where $\boldsymbol{\Phi}^{\mathrm{HR}}_r \in \mathbb{R}^{c N_p^{\mathrm{HR}} \times r}$ contains the retained spatial modes, $\boldsymbol{\Sigma}^{\mathrm{HR}}_r \in \mathbb{R}^{r \times r}$ is the corresponding diagonal matrix of singular values, and $\boldsymbol{\Psi}^{\mathrm{HR}}_r \in \mathbb{R}^{N_t \times r}$ contains the associated temporal modes.
Hereafter, $\boldsymbol{\psi}^{\mathrm{HR}}_k \in \mathbb{R}^{r}$ denotes the reduced \ac{HR} coefficient vector associated with the $k$-th snapshot, i.e. the $k$-th row of $\boldsymbol{\Psi}^{\mathrm{HR}}_r$ transposed to a column vector.

The choice of the truncation rank involves a delicate trade-off between preserving physically meaningful flow structures and rejecting noise, which typically contaminates the higher-order modes. In this sense, previous studies have proposed criteria based on the modal amplitude relative to the measurement noise \citep{epps2010}, on the degree of spatial organization retained by the modes \citep{brindise2017}, and on the progressive loss of modal reliability as noise corruption increases \citep{epps2019}. In this work, the truncation rank $r$ is selected using the \emph{elbow criterion}, restated by \citet{raiolaPIVRandomError2015a} as the relative decrease rate of the reconstruction error at the $i$-th POD mode, $F(i)$, rather than on a prescribed retained-energy threshold. The threshold $F(i)=0.999$ is adopted to determine the number of retained \ac{HR} modes. \revA{This threshold identifies the mode index beyond which adding one more mode reduces the reconstruction error by less than $0.1\%$ in relative terms, and the additional modes are expected to be dominated by measurement noise. Other values were also considered: stricter truncations discard part of the energetic broadband content of the present flows, whereas milder ones mostly add noise-dominated modes with no appreciable benefit on the estimation. The relatively high retained ranks ($r \approx 180$--$190$) reflect the broadband, multi-scale character of the two datasets, whose slowly decaying modal energy distribution (see Section~\ref{sec:results}) requires many modes to represent the resolved dynamics, rather than the specific threshold choice. Note also that the \ac{POD} truncation acts at the same time as an effective denoising stage, since measurement noise is mostly scattered across the discarded low-energy modes \citep{raiolaPIVRandomError2015a,brindise2017}.}

An analogous \ac{POD} analysis is applied to the \ac{LR} training dataset. Denoting by $\mathbf{Z}^{\mathrm{LR}} \in \mathbb{R}^{cN_p^{\mathrm{LR}} \times N_t}$ the matrix of rank $n$ collecting the \ac{LR} training snapshots, its singular value decomposition yields
\begin{equation}
    \mathbf{Z}^{\mathrm{LR}} =
    \boldsymbol{\Phi}^{\mathrm{LR}}
    \boldsymbol{\Sigma}^{\mathrm{LR}}
    \boldsymbol{\Psi}^{\mathrm{LR}^{\scriptstyle \mathsf{T}}}\,.
    \label{eq:svd_lr}
\end{equation}
In contrast to the \ac{HR} side, no truncation is applied to the \ac{LR} latent space, and the full numerical rank $n$ is retained.

During the validation and online reconstruction stage, i.e. for the snapshots belonging to the testing subset, each incoming \ac{LR} snapshot $\mathbf{z}^{\mathrm{LR}}_k$ is projected onto the previously identified full-rank \ac{LR} basis to obtain the coefficient vector
\begin{equation}
    \boldsymbol{\psi}^{\mathrm{LR}}_k =
    \boldsymbol{\Sigma}^{\mathrm{LR}^{\scriptstyle -1}}
    \boldsymbol{\Phi}^{\mathrm{LR}^{\scriptstyle \mathsf{T}}}
    \mathbf{z}^{\mathrm{LR}}_k
    \in \mathbb{R}^{n}\revA{,}
    \label{eq:lr_projection}
\end{equation}
which constitutes the common input to all three estimation methodologies described below.

In the following, we consider two-component \revA{velocity} measurements, i.e. $c=\text{2}$, although the extension to volumetric three-component data is straightforward.

\subsection{Estimation of \ac{HR} states from \ac{LR} measurements}

In this section, three estimators will be described:
\begin{itemize}
    \item \textbf{Direct Kalman filter estimation}: the \ac{HR} temporal coefficients are obtained using directly the \ac{LR} data as observables for the Kalman filter.
    \item \textbf{\ac{LSE}-based estimation with Kalman filtering}: the \ac{HR} coefficients are first estimated with an \ac{LSE} from the \ac{LR} coefficients, and then a Kalman filter is applied as a regularizer.
    \item \textbf{Variance-rescaled \ac{LSE}-based estimation}: based on the same strategy of the previous method, corrected with a mode-variance correction before the Kalman filtering to compensate for systematic attenuation of the least observable \ac{HR} modes.
\end{itemize}

The three estimators  differ in the way the reduced \ac{HR} coefficients are inferred from the \ac{LR} measurements. However, all of them rely on the same linear reduced-order state-transition model of the \ac{HR} coefficients. This common dynamical model is introduced first, and the three estimation strategies are described subsequently.

% To distinguish the snapshot matrices used in the estimation stage from the temporal mode matrices arising from the \ac{POD}, we introduce dedicated notation for the coefficient data matrices assembled snapshot-by-snapshot. Specifically, we define
% \begin{equation}
%     \begin{aligned}
%     \mathbf{A}^{\mathrm{HR}} \triangleq \boldsymbol{\Psi}^{\mathrm{HR}^{\scriptstyle \mathsf{T}}}_r \in \mathbb{R}^{r \times N_t}\,, \\
%     \mathbf{A}^{\mathrm{LR}} \triangleq \boldsymbol{\Psi}^{\mathrm{LR}^{\scriptstyle \mathsf{T}}} \in \mathbb{R}^{n \times N_t}\,,
%     \label{eq:A_def}
%     \end{aligned}
% \end{equation}
% so that the $k$-th columns of $\mathbf{A}^{\mathrm{HR}}$ and $\mathbf{A}^{\mathrm{LR}}$ coincide with the individual coefficient vectors $\boldsymbol{\psi}^{\mathrm{HR}}_k$ and $\boldsymbol{\psi}^{\mathrm{LR}}_k$, truncated at rank $r$ and $n$, respectively. Subscripts ``$\mathrm{train}$'' and ``$\mathrm{val}$'' are appended to denote the submatrices formed by retaining only the columns corresponding to the training and validation subsets. %The tilde accent $\tilde{(\cdot)}$ is used to denote estimated or rescaled counterparts of these matrices.

For the estimator identification, it is convenient to collect the reduced coordinates at all time instants as column-wise data matrices,
\begin{equation}
    \begin{aligned}
    \mathbf{A}^{\mathrm{HR}} &\equiv
    \left[
    \boldsymbol{\psi}^{\mathrm{HR}}_1,\ldots,
    \boldsymbol{\psi}^{\mathrm{HR}}_{N_t}
    \right]
    =
    \boldsymbol{\Psi}^{\mathrm{HR}\mathsf{T}}_r
    \in \mathbb{R}^{r \times N_t}, \\
    \mathbf{A}^{\mathrm{LR}} &\equiv
    \left[
    \boldsymbol{\psi}^{\mathrm{LR}}_1,\ldots,
    \boldsymbol{\psi}^{\mathrm{LR}}_{N_t}
    \right]
    =
    \boldsymbol{\Psi}^{\mathrm{LR}\mathsf{T}}
    \in \mathbb{R}^{n \times N_t}.
    \end{aligned}
\end{equation}
These matrices are introduced only as a compact notation for \revA{the state-estimation} stage: each column corresponds to one time sample of the reduced HR state or LR measurement.

\subsubsection{Common state transition model}
During online estimation, a state transition model is used to regularize the output. Considering the short prediction horizon (only one step ahead in time-resolved measurements), a linear discrete one-step model is adopted for the reduced \ac{HR} state. Specifically, the retained \ac{HR} temporal coefficients are assumed to evolve according to a linear state-transition matrix $\mathbf{F} \in \mathbb{R}^{r \times r}$, such that
\begin{equation}
    \boldsymbol{\psi}^{\mathrm{HR}}_{k+1}
    \approx
    \mathbf{F}\,\boldsymbol{\psi}^{\mathrm{HR}}_k.
\end{equation}
To estimate $\mathbf{F}$, two matrices collecting consecutive reduced-order snapshots are defined as
\begin{equation}
\begin{aligned}
    \mathbf{A}^{\mathrm{HR},-} &=
    \left[
    \boldsymbol{\psi}^{\mathrm{HR}}_1 \;\cdots\;
    \boldsymbol{\psi}^{\mathrm{HR}}_{N_t-1}
    \right] \in \mathbb{R}^{r \times (N_t - 1)}\,,\\
    \mathbf{A}^{\mathrm{HR},+} &=
    \left[
    \boldsymbol{\psi}^{\mathrm{HR}}_2 \;\cdots\;
    \boldsymbol{\psi}^{\mathrm{HR}}_{N_t}
    \right] \in \mathbb{R}^{r \times (N_t - 1)},
\end{aligned}
\end{equation}
The matrix $\mathbf{F}$ is then obtained in the least-squares sense as
\begin{equation}
    \mathbf{F}=
    \mathbf{A}^{\mathrm{HR},+}
    \left(
    \mathbf{A}^{\mathrm{HR},-}
    \right)^{\dagger}\,,
    \label{eq:F}
\end{equation}
where $(\cdot)^{\dagger}$ denotes the Moore--Penrose pseudoinverse. Therefore, $\mathbf{F}$ represents the best-fit linear evolution operator of the reduced \ac{HR} state over one time step.

\subsubsection{Method I --- Direct Kalman filter estimation}
\label{sec:method_dkf}

The first estimator directly infers the reduced \ac{HR} state from the \ac{LR} coefficient vector through a linear state-space model formulated entirely in coefficient space. Since the estimation is directly embedded in the Kalman filtering process, this method will be referred to as \ac{KF}. 

The reduced \ac{HR} state is assumed to evolve according to
\begin{equation}
    \boldsymbol{\psi}^{\mathrm{HR}}_k =
    \mathbf{F}\,\boldsymbol{\psi}^{\mathrm{HR}}_{k-1} + \mathbf{w}_k\,,
    \label{eq:dkf_dynamics}
\end{equation}
where $\mathbf{w}_k$ denotes the process noise of the Kalman-filter state model, accounting for unmodeled dynamics and mismatch between the true reduced-order evolution and its one-step linear approximation through $\mathbf{F}$. It is assumed to be zero-mean with covariance $\mathbf{Q}$, estimated offline from the residuals of the identified linear model. The measurement model relates the \ac{LR} coefficients vector to the reduced \ac{HR} state as
\begin{equation}
    \boldsymbol{\psi}^{\mathrm{LR}}_k =
    \mathbf{C}\,\boldsymbol{\psi}^{\mathrm{HR}}_k + \boldsymbol{\eta}_k\,,
    \label{eq:dkf_measurement}
\end{equation}
where $\mathbf{C} \in \mathbb{R}^{n \times r}$ is a linear measurement operator mapping \ac{HR} coefficients to \ac{LR} coefficients, and $\boldsymbol{\eta}_k$ denotes the measurement noise, assumed zero-mean with covariance $\mathbf{R}$.

The comparison between the HR prediction and the LR measurement is deliberately performed in the \ac{LR} coefficient space. Indeed, mapping a \ac{LR} measurement back to the \ac{HR} space before comparison would amount to solving an inverse problem, which is intrinsically ill-posed because the \ac{LR} data do not contain the missing high-frequency content. By contrast, projecting the predicted \ac{HR} state into the \ac{LR} space through $\mathbf{C}$ is a well-posed forward operation. The resulting mismatch is then mapped back to the \ac{HR} state through the covariance-weighted Kalman gain, ensuring a stable and consistent correction step.

The operator $\mathbf{C}$ is identified offline from the training data by ridge-regularized least squares,
\begin{equation}
    \mathbf{C} =
    \mathbf{A}^{\mathrm{LR}}_{\mathrm{train}}
    \mathbf{A}^{\mathrm{HR}^{\scriptstyle \mathsf{T}}}_{\mathrm{train}}
    \left(
        \mathbf{A}^{\mathrm{HR}}_{\mathrm{train}}
        \mathbf{A}^{\mathrm{HR}^{\scriptstyle \mathsf{T}}}_{\mathrm{train}}
        + \lambda \mathbf{I}_r
    \right)^{-1}\,,
    \label{eq:C_identification}
\end{equation}
where $\lambda$ is a small regularization parameter, here set to $\lambda = 10^{-12}$.

The process noise covariance $\mathbf{Q}$ is estimated empirically from the residuals of the one-step linear model on the training set. Specifically, the matrix
\begin{equation}
    \mathbf{E}_{\mathrm{model}} =
    \mathbf{A}^{\mathrm{HR},+} -
    \mathbf{F}\,\mathbf{A}^{\mathrm{HR},-}
\end{equation}
collects, column by column, the sample residuals of the reduced-order state equation, i.e. the observed realizations of the process-noise term $\mathbf{w}_k$ over the training set. The corresponding covariance estimate is then
\begin{equation}
    \mathbf{Q} = \frac{1}{N_t - 1}
    \mathbf{E}_{\mathrm{model}}\mathbf{E}_{\mathrm{model}}^\mathsf{T}\,.
    \label{eq:Q_dkf}
\end{equation}
Similarly, the measurement noise covariance $\mathbf{R}$ is estimated from the residuals of Eq.~\eqref{eq:dkf_measurement} over the validation subset. Denoting by $\mathbf{A}^{\mathrm{LR}}_{\mathrm{val}}$ and $\mathbf{A}^{\mathrm{HR}}_{\mathrm{val}}$ the matrices collecting the corresponding \ac{LR} and \ac{HR} coefficients vectors on that subset, one defines
\begin{equation}
    \mathbf{E}_{\mathrm{meas}} =
    \mathbf{A}^{\mathrm{LR}}_{\mathrm{val}} -
    \mathbf{C}\,\mathbf{A}^{\mathrm{HR}}_{\mathrm{val}}\,.
\end{equation}
The matrix $\mathbf{E}_{\mathrm{meas}}$ therefore collects, column by column, the sample residuals of the measurement equation, and is used as an empirical collection of realizations of the measurement-noise term $\boldsymbol{\eta}_k$. The corresponding covariance estimate is
\begin{equation}
    \mathbf{R} = \frac{1}{N_{\mathrm{val}} - 1}
    \mathbf{E}_{\mathrm{meas}}\mathbf{E}_{\mathrm{meas}}^\mathsf{T}\,.
    \label{eq:R_dkf}
\end{equation}

Once $\mathbf{F}$, $\mathbf{C}$, $\mathbf{Q}$, and $\mathbf{R}$ have been identified, the reduced \ac{HR} state is estimated online through the classical Kalman prediction-update procedure. At each k-th timestep, the Kalman gain is updated as
\begin{equation}
    \mathbf{K}_k =
    \mathbf{P}_{k|k-1}\,\mathbf{C}^\mathsf{T}
    \left(
        \mathbf{C}\,\mathbf{P}_{k|k-1}\,\mathbf{C}^\mathsf{T} + \mathbf{R}
    \right)^{-1}\,,
    \label{eq:dkf_gain}
\end{equation}
where $\mathbf{P}_{k|k-1}$ denotes the covariance matrix of the \emph{a priori} estimation error associated with the predicted reduced \ac{HR} state. Although, in principle, this quantity corresponds to the covariance of the difference between the true state and its prediction, the true state $\boldsymbol{\psi}^{\mathrm{HR}}_k$ is unknown during online estimation. Therefore, $\mathbf{P}_{k|k-1}$ is computed recursively by propagating the previous updated covariance through the state-transition model and adding the process-noise covariance, namely
\begin{equation}
    \mathbf{P}_{k|k-1}
    =
    \mathbf{F}\,\mathbf{P}_{k-1|k-1}\,\mathbf{F}^\mathsf{T}
    +
    \mathbf{Q}\,.
\end{equation}

The corrected estimate can then be written compactly as
\begin{equation}
\begin{aligned}    
    \hat{\boldsymbol{\psi}}^{\mathrm{HR}}_{k|k} =&
    \mathbf{F}\,\hat{\boldsymbol{\psi}}^{\mathrm{HR}}_{k-1|k-1}
    +\\&
    \mathbf{K}_k
    \left(
        \boldsymbol{\psi}^{\mathrm{LR}}_k
        -
        \mathbf{C}\,\mathbf{F}\,\hat{\boldsymbol{\psi}}^{\mathrm{HR}}_{k-1|k-1}
    \right)\,,
    \end{aligned}
    \label{eq:dkf_combined_update}
\end{equation}
which highlights that the updated \ac{HR} state is obtained by correcting the model-based prediction through the mismatch between the measured \ac{LR} coefficients and the corresponding projection of the predicted \ac{HR} state from the model. The associated covariance matrices are updated according to the standard Kalman filter recursion \citep{kalman1960,An_introduction_to_the_Kalman_filter}.
In the present implementation, the initial state is obtained by applying the pseudoinverse of $\mathbf{C}$ to the first available \ac{LR} measurement.

\subsubsection{Method II --- LSE-based estimation with Kalman filtering}
\label{sec:method_lse}

The second approach decouples the spatial mapping from the temporal regularization. A \ac{LSE} is first used to map the \ac{LR} coefficient vector directly onto an instantaneous estimate of the reduced \ac{HR} state. A Kalman filter is then applied only as a temporal denoising and regularization stage. This method is hereinafter referred to as \emph{LSE}.

The \ac{LSE} operator $\mathbf{M} \in \mathbb{R}^{r \times n}$ is identified offline from the training data by ridge-regularized least squares,
\begin{equation}
    \mathbf{M} =
    \mathbf{A}^{\mathrm{HR}}_{\mathrm{train}}
    \mathbf{A}^{\mathrm{LR}^{\scriptstyle \mathsf{T}}}_{\mathrm{train}}
    \left(
        \mathbf{A}^{\mathrm{LR}}_{\mathrm{train}}
        \mathbf{A}^{\mathrm{LR}^{\scriptstyle\mathsf{T}}}_{\mathrm{train}}
        + \lambda \mathbf{I}_{n}
    \right)^{-1}
    \label{eq:M_lse}
\end{equation}
where $\lambda$ is the same regularization parameter introduced in Eq.~\eqref{eq:C_identification} and set also in this case to $\lambda = 10^{-12}$. In contrast to $\mathbf{C}$ in Eq.~\eqref{eq:C_identification}, which is used in Method I as a forward operator to predict the \ac{LR} coefficients associated with a given reduced \ac{HR} state, $\mathbf{M}$ directly provides a statistical LR-to-HR mapping and is therefore used here to reconstruct an instantaneous estimate of the reduced \ac{HR} coefficients from the observed \ac{LR} ones.

At each time step, the instantaneous \ac{LSE} estimate of the reduced \ac{HR} coefficients is obtained as
\begin{equation}
    \tilde{\boldsymbol{\psi}}^{\mathrm{HR}}_k =
    \mathbf{M}\,\boldsymbol{\psi}^{\mathrm{LR}}_k\,.
    \label{eq:lse_estimate}
\end{equation}
This estimate is optimal in the regularized least-squares sense over the training data, but it remains affected by frame-to-frame noise inherited from the \ac{LR} measurement and tends to attenuate the variance of the least observable higher-order modes.

The Kalman filter is then introduced only as a temporal filtering stage acting on the \ac{LSE}-based estimate. The reduced \ac{HR} state is assumed to evolve according to
\begin{equation}
    \boldsymbol{\psi}^{\mathrm{HR}}_k =
    \mathbf{F}\,\boldsymbol{\psi}^{\mathrm{HR}}_{k-1} + \mathbf{w}_k\,,
    \label{eq:lse_dynamics}
\end{equation}
while the measurement model is written directly in the \ac{HR} coefficient space as
\begin{equation}
    \tilde{\boldsymbol{\psi}}^{\mathrm{HR}}_k =
    \boldsymbol{\psi}^{\mathrm{HR}}_k + \boldsymbol{\eta}_k\,,
    \label{eq:lse_measurement}
\end{equation}
where $\boldsymbol{\eta}_k$ is a zero-mean noise term with covariance $\mathbf{R}_{\mathrm{LSE}}$.

Because the \ac{LSE} output already provides an estimate in the reduced \ac{HR} space, no additional measurement operator that provides a mapping between the LR and HR space is needed, and the Kalman filter acts directly on the mismatch between the model prediction and the \ac{LSE}-based estimate. The Kalman gain, therefore, simplifies to
\begin{equation}
    \mathbf{K}_k =
    \mathbf{P}_{k|k-1}
    \left(
        \mathbf{P}_{k|k-1} + \mathbf{R}_{\mathrm{LSE}}
    \right)^{-1}
    \label{eq:lse_gain}\,,
\end{equation}
where $\mathbf{P}_{k|k-1}$ denotes the predicted state-error covariance matrix. The corrected estimate can then be written as
\begin{equation}
\begin{aligned}    
    \hat{\boldsymbol{\psi}}^{\mathrm{HR}}_{k|k} =&
    \mathbf{F}\,\hat{\boldsymbol{\psi}}^{\mathrm{HR}}_{k-1|k-1}
    +\\&
    \mathbf{K}_k
    \left(
        \mathbf{M}\,\boldsymbol{\psi}^{\mathrm{LR}}_k\,
        -
        \mathbf{F}\,\hat{\boldsymbol{\psi}}^{\mathrm{HR}}_{k-1|k-1}
    \right)\,,
    \end{aligned}
    \label{eq:lse_update_state}
\end{equation}
which highlights that the updated state is obtained by correcting the model prediction through the discrepancy between the instantaneous \ac{LSE} estimate and the predicted reduced \ac{HR} state. The associated covariance matrices are updated according to the standard Kalman filter recursion.

The process noise covariance is again estimated from Eq.~\eqref{eq:Q_dkf}, whereas the measurement noise covariance is estimated from the error between the \ac{LSE} estimate and the reference \ac{HR} coefficients on the validation set,
\begin{equation}
\begin{aligned}
    \mathbf{R}_{\mathrm{LSE}} &= \frac{1}{N_{\mathrm{val}}-1}
    \mathbf{E}_{\mathrm{LSE}}\mathbf{E}_{\mathrm{LSE}}^\mathsf{T} \,,\\
    \mathbf{E}_{\mathrm{LSE}} &=
    \tilde{\mathbf{A}}^{\mathrm{HR}}_{\mathrm{val}} -
    \mathbf{A}^{\mathrm{HR}}_{\mathrm{val}}\,.
    \label{eq:R_lse}
\end{aligned}
\end{equation}

\subsubsection{Method III --- Variance-rescaled \ac{LSE}-based estimation}
\label{sec:method_vr_lse}

The third method extends Method II by introducing a heuristic, mode-wise variance correction during the estimation stage. Its purpose is to compensate for the systematic attenuation of the least observable higher-order \ac{HR} modes, which is a known limitation of regression-based estimators \citep{ewing1999,tinney2006}. The correction is identified offline on the training set and then applied during the online stage.
This method is hereinafter referred to as \emph{LSE+VR}.

A diagonal rescaling matrix $\boldsymbol{\Gamma} \in \mathbb{R}^{r \times r}$ is introduced as
\begin{equation}
    \boldsymbol{\Gamma} = \text{diag}\left(
    \left(
        \mathbf{A}^{\mathrm{HR}}_{\mathrm{train}}
        \mathbf{A}^{\mathrm{HR}^{\scriptstyle \mathsf{T}}}_{\mathrm{train}}
    \right)\oslash\left(
        \tilde{\mathbf{A}}^{\mathrm{HR}}_{\mathrm{train}}
        \tilde{\mathbf{A}}^{\mathrm{HR}^{\scriptstyle \mathsf{T}}}_{\mathrm{train}}
    \right)\right)^{\circ \frac{1}{2}}
    \,,
    \label{eq:var_rescaling_coeff}
\end{equation}

where $\text{diag}(\cdot)$ retains only the diagonal entries of its argument as a diagonal matrix, $\oslash$ indicates the Hadamard elementwise division, and the $\circ \frac{1}{2}$ superscript indicates the elementwise Hadamard square root.
Each diagonal entry of $\boldsymbol{\Gamma}$ then equals the ratio of the empirical standard deviation of the corresponding true \ac{HR} coefficients to that of its \ac{LSE} estimate over the training set: multiplying the prediction by $\boldsymbol{\Gamma}$ restores the correct variance level on average, hence the name \textit{variance rescaling}.

Accordingly, the intermediate estimate entering the Kalman correction is redefined as
\begin{equation}
    \tilde{\boldsymbol{\psi}}^{\mathrm{HR}}_k =
    \boldsymbol{\Gamma}\,\mathbf{M}\,\boldsymbol{\psi}^{\mathrm{LR}}_k\,.
    \label{eq:lse_vr}
\end{equation}
Method III therefore retains the same dynamical model, Kalman gain, and state-update equations introduced in Method II, the only difference being the use of the variance-rescaled intermediate estimate in place of the standard \ac{LSE} prediction.

The process noise covariance is again estimated from Eq.~\eqref{eq:Q_dkf}, whereas the measurement noise covariance is obtained from the error between the variance-rescaled estimates and the reference \ac{HR} coefficients on the validation set,
\begin{equation}
\begin{aligned}
    \mathbf{R}_{\mathrm{VR}} &= \frac{1}{N_{\mathrm{val}}-1}
    \mathbf{E}_{\mathrm{VR}}\mathbf{E}_{\mathrm{VR}}^\mathsf{T} \,,\\
    \mathbf{E}_{\mathrm{VR}} &=
    \tilde{\mathbf{A}}^{\mathrm{HR}}_{\mathrm{val}} -
    \mathbf{A}^{\mathrm{HR}}_{\mathrm{val}}\,.
    \label{eq:R_vr}
\end{aligned}
\end{equation}

\subsection{\ac{HR} field reconstruction and baseline comparison}
\label{sec:reconstruction}

For all three methods, once the \revC{Kalman-filtered estimate $\hat{\boldsymbol{\psi}}^{\mathrm{HR}}_{k|k}$} of the reduced \ac{HR} coefficient vector is available, the corresponding fluctuating \ac{HR} snapshot is reconstructed through the retained \ac{HR} POD basis as
\begin{equation}
    \hat{\mathbf{z}}^{\mathrm{HR}}_k =
    \boldsymbol{\Phi}^{\mathrm{HR}}_r \boldsymbol{\Sigma}^{\mathrm{HR}}_r
    \revC{\hat{\boldsymbol{\psi}}^{\mathrm{HR}}_{k|k}}\,,
    \label{eq:reconstruction}
\end{equation}
\revC{Throughout the paper, the tilde accent $\tilde{(\cdot)}$ denotes instantaneous (pre-filtering) estimates, such as the \ac{LSE} output of Eq.~\eqref{eq:lse_estimate}, whereas the hat accent $\hat{(\cdot)}$ denotes the Kalman-filtered estimates; the field reconstruction is always based on the latter.}
The full \ac{HR} velocity field is then recovered by reshaping \revC{$\hat{\mathbf{z}}^{\mathrm{HR}}_k$} onto the \ac{HR} grid and adding $\overline{\mathbf{U}}^{\mathrm{HR}}$.

The proposed methods are compared against direct cubic interpolation of the \ac{LR} velocity field onto the \ac{HR} grid. The choice is driven by the need to perform the spatial upsampling in real-time, thus preventing the use of more advanced super-resolution or data-assimilation strategies. Unlike the reduced-order estimators introduced above, cubic interpolation does not exploit either the temporal evolution of the flow or the cross-resolution relationships learned from the paired \ac{HR}/\ac{LR} training data. It is therefore adopted as a reference baseline to assess the benefit of the proposed data-driven approaches.

% ============================================================
%  SECTION 3 — Experimental Datasets
% ============================================================
\section{Experimental datasets and processing pipeline}
\label{sec:exp_setup}

The estimation framework introduced in Section~\ref{sec:methodology} is evaluated on two experimental turbulent flow datasets acquired via pulsed-\ac{EBIV}: a submerged axisymmetric water jet and an air channel flow obstructed by a spanwise square rib. These two cases were selected as representative of broadband flows, in which energy is distributed over a wide range of spatial and temporal scales. This makes them a challenging testbed for reduced-order estimation and cross-resolution reconstruction.

\begin{figure*}
    \centering
    \includegraphics{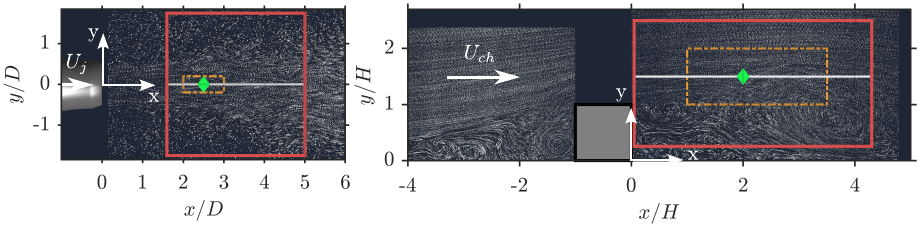}
    \caption{Representative multi-exposed \ac{EBV} pseudo-snapshots of the two experimental configurations considered in the present work: submerged round jet (left) and channel flow over a square rib (right). The images are formed by accumulating four consecutive pulses for the jet and ten consecutive pulses for the rib case, in order to enhance the visualization of the flow structures. The so constructed images are not used for the data processing described in Sec~\ref{sec:ebiv_proc}. The red rectangle marks the resolved region retained for the present study; the orange dashed rectangle identifies the subdomain used for the temporal premultiplied \acs{PSD} analysis; the green marker denotes the probe location used for the autocorrelation analysis; the white horizontal line indicates the sampling line adopted for the spatial \acs{PSD}. For detailed sketches of the full experimental setups, the reader is referred to \citet{franceschelliAssessmentEventbasedImaging2025}.}
    \label{fig:exp_domains}
\end{figure*}

The pulsed-\ac{EBIV} principle adopted here follows the approach introduced by \citet{willertEventbasedImagingVelocimetry2023}. The channel flow dataset is based on the experimental campaigns documented in \citet{franceschelliAssessmentEventbasedImaging2025}, while the jet flow one is obtained from an upgraded facility compared to the one described in the same work, but with similar optical arrangement. Both the setups are summarized here for completeness, together with the specific processing choices adopted in the present work.
In both configurations, the coordinate system shown in Figure~\ref{fig:exp_domains} is adopted, with $x$ and $y$ denoting the streamwise and transverse directions, respectively.
The main parameters of the two experimental configurations and of the corresponding \ac{LR} and \ac{HR} processing pipelines are summarized in Table~\ref{tab:setup}. 

\begin{table*}[ht]
  \centering
  \caption{Summary of experimental and processing parameters for
           the two flow configurations.}
  \label{tab:setup}
  \renewcommand{\arraystretch}{1.25}
  \begin{tabular}{lcc}
    \hline
    Parameter & Jet flow & \revA{Channel flow} \\
    \hline
    Medium                             & Water          & Air             \\
    Characteristic length              & $D = 30\,\mathrm{mm}$
                                       & $H = 8.18\,\mathrm{mm}$         \\
    Reference velocity $U_\mathrm{ref}$& $0.2\,\mathrm{m/s}$
                                       & $2.8\,\mathrm{m/s}$             \\
    Reynolds number                    & $Re_D \approx 6\,000$
                                       & $Re_H \approx 1\,500$           \\
    Acquisition rate                   & $100\,\mathrm{Hz}$
                                       & $5\,\mathrm{kHz}$               \\
    Laser pulse width                  & \revC{$1\,\mathrm{ms}$}
                                       & $< 50\,\mathrm{ns}$             \\
    EBV camera                         & \multicolumn{2}{c}{Prophesee EVK4
                                         (Sony IMX636)}                  \\
    Spatial resolution [pixel/mm]      & $\approx 6.5$  & $\approx 30.3$  \\
    Pixel displacement corresponding to $U_{ref}$ [pixel/pseudo-frame]
                                       & $\approx 13.0$ & $\approx 17.0$  \\
    \hline
    \multicolumn{3}{l}{\textit{LR processing (real-time pipeline)}}      \\
    \hline
    Algorithm                          & \multicolumn{2}{c}{Single-pass
                                         cross-correlation}              \\
    IW size [pixel]                    & $48\times48$ & $64\times64$ \\
    Overlap                            & \multicolumn{2}{c}{$0\%$}       \\
    Window deformation                 & \multicolumn{2}{c}{No}           \\
    \hline
    \multicolumn{3}{l}{\textit{HR processing (offline reference)}}       \\
    \hline
    Algorithm                          & \multicolumn{2}{c}{\revA{Multi-frame,
                                         multi-pass cross-correlation}}       \\
    Temporal stencil                   & 3 frames & 5 frames     \\
    IW size [pixel]                    & $32\times32$ & $64\times64$ \\
    Overlap                            & \multicolumn{2}{c}{$75\%$}       \\
    Window deformation                 & \multicolumn{2}{c}{Yes}          \\
    HR/LR vector-spacing ratio         & $6$ & $4$          \\
    \hline
    Total snapshots $N_s$              & \multicolumn{2}{c}{$9\,000$}     \\
    \hline
  \end{tabular}
\end{table*}

% ------------------------------------------------------------
\subsection{Submerged water jet}
\label{sec:setup_jet}

The first dataset corresponds to a submerged round jet issuing horizontally into a water tank of dimensions $900 \times 500 \times 450$~mm$^3$, corresponding to the tank length, width, and height, aligned with the $x$-, $z$-, and $y$-directions, respectively. The nozzle exit diameter is $D = 30$ mm, and the bulk exit velocity is approximately $U_j = 0.2$ m/s, yielding a Reynolds number of $\mathrm{Re}_D = U_j D / \nu \approx 6000$, with $\nu$ being the kinematic viscosity. The tank is seeded with polyamide tracer particles of nominal diameter $56\,\mu$m.

The measurement plane, passing through the jet axis,  is illuminated by a pulsed laser sheet generated by a low-cost continuous-wave diode laser originally designed for engraving applications (LaserTree LT-40W-AA, \revA{with $5\,\mathrm{W}$ optical power}), operated in pulse-width-modulated mode. The light sheet thickness is below 2 mm at the measurement location\revA{, as estimated by imaging the sheet cross-section on a calibration target during the optical alignment}. The pulse repetition rate is $f_{j}=100$ Hz, with a pulse width of 1 ms.

Data are recorded using a Prophesee EVK4 event-based camera equipped with a Sony IMX636 sensor featuring a resolution of $1280 \times 720$ pixels and a pixel pitch of $4.86\,\mu\mathrm{m}$. The optical magnification yields a spatial resolution of approximately 6.5 pixel/mm over a field of view of about $190 \times 110$ mm$^2$. The resolved area considered in the present study spans $1.6 \leq x/D \leq 5$ in the streamwise direction and $-1.75 \leq y/D \leq 1.75$ in the radial direction. The acquired dataset consists of 9000 velocity snapshots.

A representative multi-exposed \ac{EBV} pseudo-snapshot of the jet configuration is shown in Figure~\ref{fig:exp_domains}. For visual clarity, the image is formed here by combining four consecutive laser pulses. The figure also reports the spatial domains and sampling locations used in the analyses presented later in this work, including the resolved region, the subdomain adopted for the temporal premultiplied \ac{PSD}, the probe location used for the autocorrelation analysis, and the sampling line used for the spatial \ac{PSD}. For a detailed sketch of the complete experimental setup, the reader is referred to \citet{franceschelliAssessmentEventbasedImaging2025}.

% ------------------------------------------------------------
\subsection{Channel flow over a square rib}
\label{sec:setup_rib}

The channel-flow dataset is derived from the experimental campaign described by \citet{franceschelliAssessmentEventbasedImaging2025}, to which the reader is referred for full details of the facility, optical arrangement, seeding system, and camera calibration procedure. Briefly, a spanwise square rib of \revA{height} $H = 8.18\,\mathrm{mm}$ is mounted on the lower wall of a square-section channel of width $W = 76\,\mathrm{mm}$. The free-stream velocity is $U_{ch} \approx 2.8\,\mathrm{m/s}$, corresponding to a rib Reynolds number of $\mathrm{Re}_H \approx 1500$. The flow is seeded with paraffin droplets of nominal diameter $1\,\mu\mathrm{m}$, injected into the upstream settling chamber, and illuminated by a high-speed pulsed Nd:YVO4 laser (Innolas/Iradion Nanio Air 532-10-V-SP). The laser pulse duration is $\sim 50\,\mathrm{ns}$ and the repetition rate is $f_{ch} = 5\,\mathrm{kHz}$. \ac{EBV} data are acquired using the same Prophesee EVK4 camera employed for the jet experiment, with a spatial resolution of $30.3\,\mathrm{pixel/mm}$ over a field of view of approximately $42\,\mathrm{mm} \times 24\,\mathrm{mm}$, corresponding to about $5H \times 3H$.

In the present work, only the wake region downstream of the rib is considered, covering $0 \leq X/H \leq 4.25$ in the streamwise direction and $0.25 \leq Y/H \leq 2.25$ in the wall-normal direction, with the origin located at the intersection between the channel wall and the rear-facing side of the rib. This region includes \revA{the separated shear layer,} the recirculation bubble, the reattachment zone, and the initial downstream recovery, thus encompassing the most dynamically active portion of the flow.

The complete dataset contains $30\,000$ velocity snapshots. In the present study, a contiguous subset of $9\,000$ snapshots is retained to ensure consistency with the jet dataset.

A representative multi-exposed \ac{EBV} pseudo-snapshot of the rib configuration is also reported in Figure~\ref{fig:exp_domains}. In this case, the visualization is formed by combining ten consecutive laser pulses in order to better highlight the wake structure. As for the jet case, the figure identifies the resolved domain retained for the present study together with the subdomain and sampling locations used for the spectral and correlation analyses discussed in Section~\ref{sec:results}.
% [R1] Removed repeated final sentence referring the reader again to Franceschelli et al. (2025).

% ------------------------------------------------------------
\subsection{\ac{EBV} camera data processing}
\label{sec:setup_ebiv}

The raw asynchronous event stream is partitioned into pseudo-snapshots following the pulsed-\ac{EBIV} procedure introduced by \citet{willertEventbasedImagingVelocimetry2023}. Each laser pulse generates a burst of positive-polarity events as the illuminated tracer particles cross the per-pixel contrast threshold of the \ac{EBV} sensor. The laser trigger and the event stream are synchronized through a common TTL signal, allowing the accumulation of all positive events associated with a given laser pulse over a user-defined accumulation time. A pseudo-image is therefore obtained by accumulating all positive events registered within a prescribed temporal window starting from the onset of each laser pulse. The resulting images are binary: pixels that register at least one event are assigned a value of unity, whereas pixels with no activation are assigned zero.

For both experimental cases, the binary pseudo-images are subsequently smoothed with a Gaussian kernel of standard deviation $0.75\,\mathrm{pixel}$ prior to cross-correlation processing. In addition, the internal bias settings of the EVK4 camera were adjusted to suppress negative-polarity events, thereby maximizing the available bandwidth for positive events and reducing the risk of arbiter saturation.

% ------------------------------------------------------------
\subsection{Low- and high-resolution \ac{EBIV} processing}
\label{sec:ebiv_proc}

The same sequence of \ac{EBIV} pseudo-snapshots is processed in two different ways: a computationally lightweight \ac{LR} pipeline representative of a realistic real-time implementation, and a more accurate \ac{HR} pipeline employed offline to generate the reference data used for training and validation.

The \ac{LR} velocity fields are obtained by single-pass cross-correlation with a fixed interrogation window, without window deformation or iterative refinement. Since only one correlation step is performed for each window pair, this procedure is representative of the computational budget available in a real-time implementation. No additional post-processing is applied apart from \revA{the universal outlier detection \citep{westerweelUniversalOutlier2005}}.

The \ac{HR} velocity fields are instead computed using an in-house multi-frame, multi-pass cross-correlation algorithm. At each time step, a stencil of consecutive pseudo-snapshots is used to improve the quality of the correlation peak, which is particularly beneficial for \ac{EBIV} data, where individual pseudo-images are binary and contain limited intensity information. The resulting \ac{HR} fields provide a spatially refined and temporally regularized reference that the estimator is trained to recover from the corresponding \ac{LR} measurements.

Table \ref{tab:setup} reports the interrogation-window sizes, overlaps and temporal stencil dimensions adopted for each dataset in both processing ways.

\revC{A remark is due on the different interrogation-window overlap adopted in the two pipelines ($0\%$ for \ac{LR}, $75\%$ for \ac{HR}). Overlapping windows increase the vector density but not, per se, the spatial resolution: cross-correlation over a window of linear size $L_{IW}$ acts as a spatial moving-average filter on the underlying displacement field, with a frequency response set by the window size and by the selected weighting window \citep{astarita2007analysis}. Velocity fluctuations of streamwise wavelength $\lambda_x$ are thus increasingly attenuated as $\lambda_x$ approaches $L_{IW}$, and can be considered well resolved only for $\lambda_x \gtrsim 2 L_{IW}$; moreover, neighbouring \ac{HR} vectors share particle-image information and exhibit a processing-induced spatial correlation over distances of the order of $L_{IW}$, regardless of the overlap. For the present \ac{HR} processing, this bound corresponds to well-resolved content for $D/\lambda_x \lesssim 3$ in the jet case and $H/\lambda_x \lesssim 2$ in the channel-flow case. The actual role of the $75\%$ overlap is to sample this window-filtered field without aliasing: at $0\%$ overlap the sampling spacing equals the filter size itself, so that the attenuated but non-negligible content in the band $L_{IW} \lesssim \lambda_x \lesssim 2L_{IW}$ is aliased, whereas the fourfold oversampling represents it cleanly. The genuine gains of the \ac{HR} pipeline over the \ac{LR} one come instead from a combination of the smaller interrogation window (although not strictly required), the window deformation, which improves the spectral response toward the filter bound \citep{scarano2002,astarita2007analysis}, and the multi-frame temporal stencil, which improves the correlation signal-to-noise ratio. Notably, for the channel-flow case the two pipelines share the same window size, so that the \ac{HR} reference mainly provides denser, alias-free sampling and higher fidelity rather than a wider resolved-wavenumber range. The learned \ac{LR}-to-\ac{HR} mapping therefore targets the \ac{HR} representation as produced by the offline pipeline, including its correlation structure: the estimators recover physically independent fine-scale content only up to the effective spatial resolution of the \ac{HR} processing, and reproduce its smoothing characteristics beyond that limit. This should be kept in mind when interpreting the spectral analyses of Section~\ref{sec:results_spectra}, where the recovered small-scale content is bounded by the resolution of the \ac{HR} reference itself.}

The datasets are then divided into training, validation, and testing subsets. The training subset is used to construct the reduced-order spaces and identify the mapping operators, the validation subset is used to estimate the Kalman-filter noise statistics (Section~\ref{sec:method_dkf}), and the testing subset is reserved for the final performance assessment. The dataset is split into $4\,500$ snapshots for training, $1\,500$ for validation, and $2\,000$ for testing, with two buffer regions of $500$ snapshots inserted between consecutive subsets to reduce temporal correlation.

%\FloatBarrier
% ============================================================
%  SECTION 4 — Results
% ============================================================
\section{Results}
\label{sec:results}

\begin{figure}
    \centering
    \includegraphics{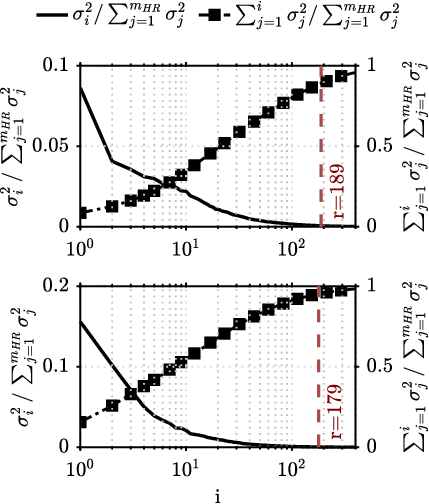}
    \caption{\revA{\ac{POD} modal energy spectrum and cumulative \ac{POD} energy} for the two considered HR datasets: top, jet flow; bottom, channel flow. The solid black line refers to the left vertical axis and reports the normalized contribution of the $i$-th mode, $\sigma_i^2 / \sum_{j=1}^{m} \sigma_j^2$. The black dashed line with square markers refers to the right vertical axis and reports the corresponding cumulative contribution, $\sum_{j=1}^{i} \sigma_j^2 / \sum_{j=1}^{m} \sigma_j^2$. The red vertical dashed line indicates the truncation rank $r$ selected by the elbow criterion.}
    \label{fig:sigma_energy}
\end{figure}

\begin{figure*}[tp]%[h]
    \centering
    \begin{tikzpicture}
        \node[anchor=south west, inner sep=0] (img1) at (0,0)
            {\includegraphics{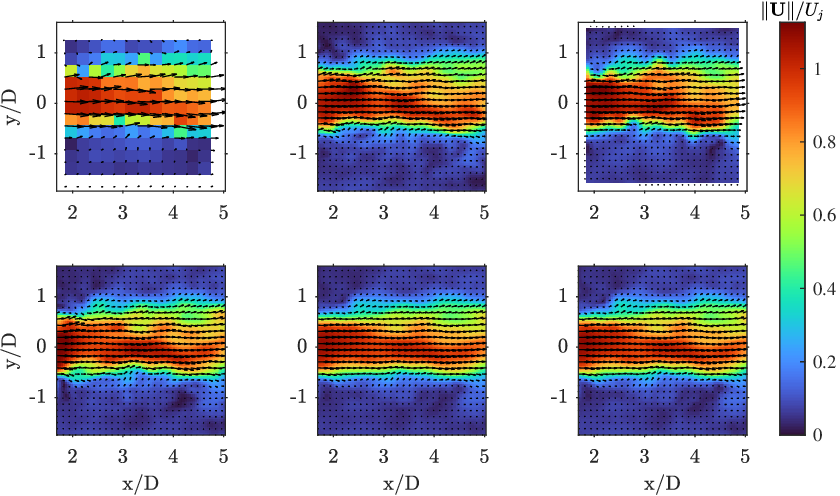}};

        % Approximate panel labels for Figure 1
        \node[anchor=north west] at ([xshift=0.8cm,yshift=0.25cm]img1.north west) {\textbf{a)} \revA{LR}};
        \node[anchor=north west] at ([xshift=5.2cm,yshift=0.25cm]img1.north west) {\textbf{b)} \revA{HR-LOR}};
        \node[anchor=north west] at ([xshift=9.7cm,yshift=0.25cm]img1.north west) {\textbf{c)} \revA{Cubic}};

        \node[anchor=north west] at ([xshift=0.8cm,yshift=-3.9cm]img1.north west) {\textbf{d)} \revA{KF}};
        \node[anchor=north west] at ([xshift=5.2cm,yshift=-3.9cm]img1.north west) {\textbf{e)} \revA{LSE}};
        \node[anchor=north west] at ([xshift=9.7cm,yshift=-3.9cm]img1.north west) {\textbf{f)} \revA{LSE+VR}};
    \end{tikzpicture}
    \caption{Comparison of instantaneous velocity-field realizations for the jet-flow case. Panels show: (a) input LR field; (b) reference HR-LOR field; (c) HR field obtained by direct cubic interpolation of the LR measurement; (d) HR estimate obtained with KF; (e) HR estimate obtained with LSE; and (f) HR estimate obtained with LSE+VR. Color contours represent the instantaneous velocity magnitude $\|\mathbf{U}\|$, normalized by the jet bulk velocity $U_j$, while arrows indicate the in-plane velocity vectors. For readability, one vector every three is shown in the HR panels.}
    \label{fig:SnapComparison_jet}
% \end{figure*}

% \begin{figure*}[h]
    \centering
    \begin{tikzpicture}
        \node[anchor=south west, inner sep=0] (img2) at (0,0)
            {\includegraphics{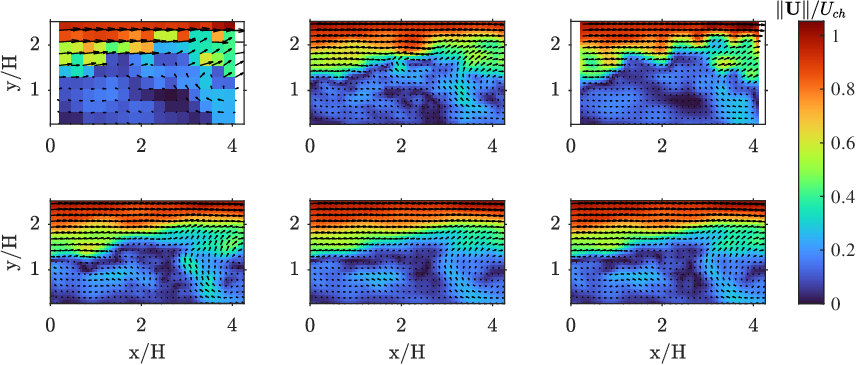}};

        % Approximate panel labels for Figure 2
        \node[anchor=north west] at ([xshift=0.7cm,yshift=0.25cm]img2.north west) {\textbf{a)} \revA{LR}};
        \node[anchor=north west] at ([xshift=5.1cm,yshift=0.25cm]img2.north west) {\textbf{b)} \revA{HR-LOR}};
        \node[anchor=north west] at ([xshift=9.6cm,yshift=0.25cm]img2.north west) {\textbf{c)} \revA{Cubic}};

        \node[anchor=north west] at ([xshift=0.7cm,yshift=-2.8cm]img2.north west) {\textbf{d)} \revA{KF}};
        \node[anchor=north west] at ([xshift=5.1cm,yshift=-2.8cm]img2.north west) {\textbf{e)} \revA{LSE}};
        \node[anchor=north west] at ([xshift=9.6cm,yshift=-2.8cm]img2.north west) {\textbf{f)} \revA{LSE+VR}};
    \end{tikzpicture}
    \caption{Comparison of instantaneous velocity-field realizations for the channel-flow case. Panels show: (a) input LR field; (b) reference HR-LOR field; (c) HR field obtained by direct cubic interpolation of the LR measurement; (d) HR estimate obtained with KF; (e) HR estimate obtained with LSE; and (f) HR estimate obtained with LSE+VR. Color contours represent the instantaneous velocity magnitude $\|\mathbf{U}\|$, normalized by the free-stream channel flow velocity $U_{ch}$, while arrows indicate the in-plane velocity vectors. For readability, one vector every two is shown in the HR panels.}
    \label{fig:SnapComparison_ch}
\end{figure*}

This section evaluates the three proposed estimation strategies on the jet and channel-flow datasets. The discussion is organized hierarchically, from instantaneous field reconstruction to increasingly demanding statistical and spectral diagnostics. Throughout the section, the reference truth used for quality metrics is assumed to be the \revA{\ac{HR} low-order reconstruction} (referred in short as \ac{HR}-\ac{LOR}), obtained by projecting the \ac{HR} dataset onto the truncated \ac{POD} basis introduced in Section~\ref{sec:methodology}. The three estimators (\ac{KF}, \ac{LSE}, and \ac{LSE}+\ac{VR}) are compared against this reference and against direct cubic interpolation of the \ac{LR} field onto the \ac{HR} grid, which is adopted as a purely spatial, non-model-based baseline.

\revA{It must be stressed that the \ac{HR}-\ac{LOR} reference is not an independent ground truth. Rather, it is the low-order projection of the \ac{HR} fields obtained from the same event stream through the offline processing pipeline. The reported errors therefore quantify how closely the real-time estimators recover the offline-grade reduced-order representation, rather than their absolute deviation from the true flow state. This is consistent with the objective of the present framework, which is precisely to make an offline-grade representation available at real-time rates. The fidelity of this target with respect to conventional velocimetry has, moreover, already been established: in \citet{franceschelliAssessmentEventbasedImaging2025}, pulsed-\ac{EBIV} measurements on the same flow configurations were benchmarked against synchronized, simultaneous \ac{PIV} measurements ---  which can fairly be regarded as the reference --- with the low-order reconstructions deviating from their \ac{PIV} counterparts by errors of the order of $0.5\%$ of the peak turbulent kinetic energy for the jet flow, and a few times higher for the channel flow at the truncation ranks of interest here.}

\subsection{\revA{POD modal energy distribution and retained energy}}

The resulting \revA{\ac{POD} modal energy distributions} for the two \ac{HR} datasets are reported in Figure~\ref{fig:sigma_energy}. \revA{Since the energy captured by the $i$-th \ac{POD} mode within the dataset equals the square of the corresponding singular value, $\sigma_i^2$, the singular-value distribution coincides with the \ac{POD} modal energy spectrum, and its running sum with the cumulative retained energy.} For each case, the figure shows the normalized modal energy  $\sigma_i^2$ together with its cumulative contribution as functions of the mode index $i$. The red vertical line indicates the truncation rank selected by the \textit{elbow criterion} introduced above. The adopted threshold yields $r=189$ for the jet case and $r=179$ for the channel-flow case, corresponding to retained cumulative energies of 89.1\% and 95.2\%, respectively. In both datasets, the singular-value decay is gradual rather than characterized by a sharp cut-off. As a result, the selected truncation lies in the high-order portion of the spectrum while still retaining most of the resolved modal energy. This behavior highlights the broadband character of the two flow cases and the wide distribution of energy across spatial scales, making these datasets particularly challenging for reduced-order reconstruction and estimation.

\subsection{Instantaneous field reconstruction}
\label{sec:results_inst}
A first qualitative assessment is presented in Fig.~\ref{fig:SnapComparison_jet} and Fig.~\ref{fig:SnapComparison_ch} for the jet and channel-flow cases, respectively, depicting the instantaneous velocity fields.
Each figure compares the input \ac{LR} field (a), the \ac{HR}-\ac{LOR} reference (b), the field obtained by direct cubic interpolation (c), and the three reconstructed fields obtained with \ac{KF} (d), \ac{LSE} (e), and \ac{LSE}+\ac{VR} (f). The color contours represent the instantaneous velocity magnitude $\|\mathbf{U}\|$, while the quiver plot indicates the in-plane velocity vector. The \ac{LR} field provides only a coarse and noisy representation of the instantaneous flow, whereas the \ac{HR}-\ac{LOR} field reveals substantially finer spatial detail, including sharp velocity gradients and the coherent structures that characterize each flow: the shear layers and vortical roll-up of the jet core, and the separated region downstream of the square rib in the channel-flow case. All three proposed estimators recover the broad spatial organization of the \ac{HR}-\ac{LOR} field with good fidelity. The \ac{KF} estimate provides a smooth reconstruction that closely follows the large-scale coherent features of the reference. The \ac{LSE} reconstruction captures the same dominant structures, although the smallest-scale features appear less sharply defined. The \ac{LSE}+\ac{VR} reconstruction recovers additional small-scale content while maintaining spatial regularity. In contrast, the cubic-interpolation baseline introduces visible high-frequency oscillations and artificial spatial modulations that are absent in the reference field. These artifacts are particularly evident in the channel-flow case, where the interpolation of the noisy \ac{LR} measurements produces spurious patterns in regions of strong velocity gradient.

\subsection{Statistics of the reconstructed fields}
\label{sec:results_stats}

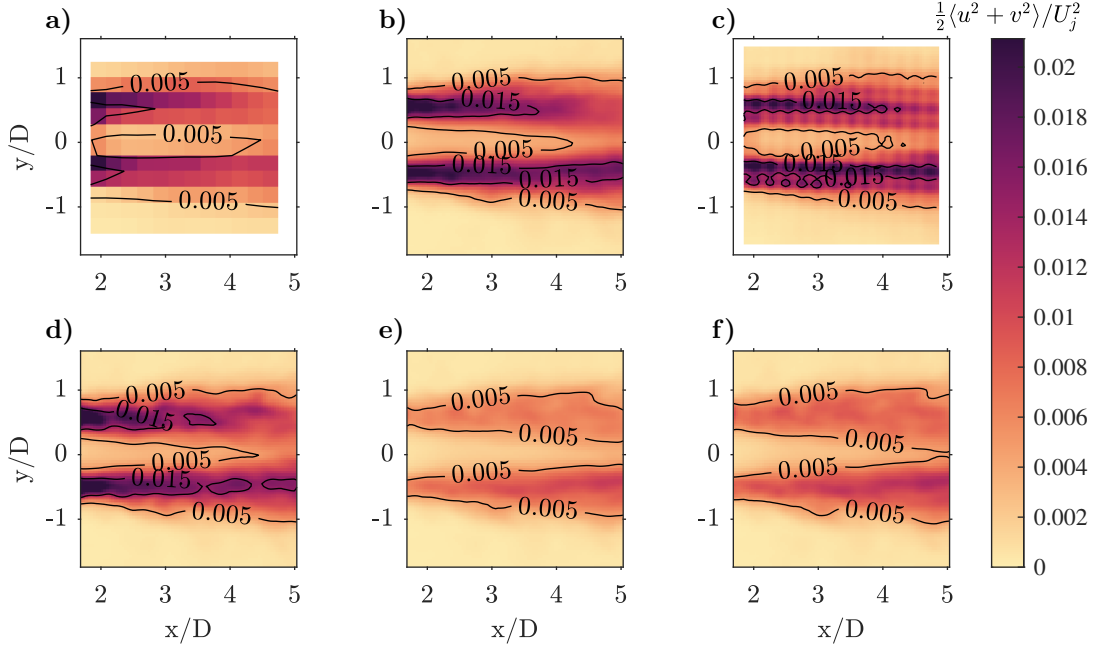
\begin{figure*}[t]
    \centering
    \begin{tikzpicture}
        \node[anchor=south west, inner sep=0] (img1) at (0,0)
            {\includegraphics{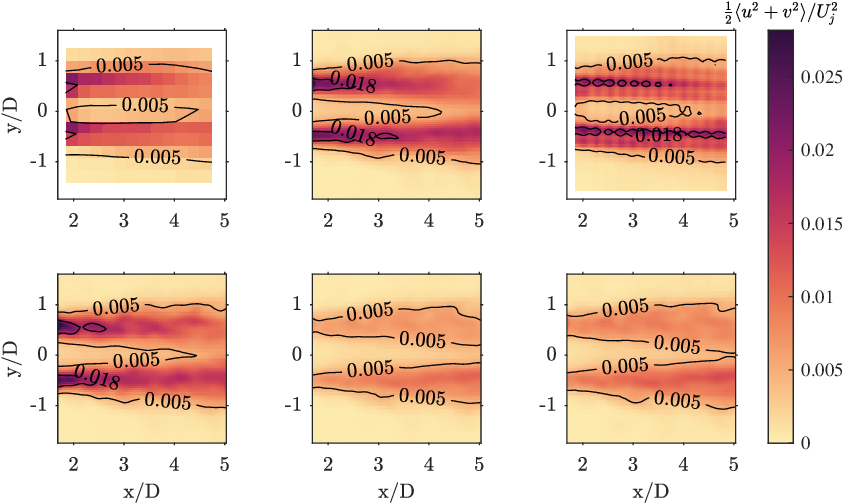}};

        % Approximate panel labels for Figure 1
        \node[anchor=north west] at ([xshift=0.85cm,yshift=0.1cm]img1.north west) {\textbf{a)} \revA{LR}};
        \node[anchor=north west] at ([xshift=5.1cm,yshift=0.1cm]img1.north west) {\textbf{b)} \revA{HR-LOR}};
        \node[anchor=north west] at ([xshift=9.5cm,yshift=0.1cm]img1.north west) {\textbf{c)} \revA{Cubic}};

        \node[anchor=north west] at ([xshift=0.85cm,yshift=-4.05cm]img1.north west) {\textbf{d)} \revA{KF}};
        \node[anchor=north west] at ([xshift=5.1cm,yshift=-4.05cm]img1.north west) {\textbf{e)} \revA{LSE}};
        \node[anchor=north west] at ([xshift=9.5cm,yshift=-4.05cm]img1.north west) {\textbf{f)} \revA{LSE+VR}};
    \end{tikzpicture}
   \caption{\revA{Spatial distribution of \ac{TKE} for the jet-flow case, normalized by the square of the jet bulk velocity $U_{j}$. Panels show: (a) \ac{LR} input; (b) \ac{HR}-\ac{LOR} reference; (c) cubic interpolation; (d) \ac{KF}; (e) \ac{LSE}; and (f) \ac{LSE}+\ac{VR}. Black isolines denote selected \ac{TKE} levels.}}
   \label{fig:tke_jet}
\end{figure*}

\begin{figure*}[t]
    \centering
    \begin{tikzpicture}
        \node[anchor=south west, inner sep=0] (img2) at (0,0)
            {\includegraphics{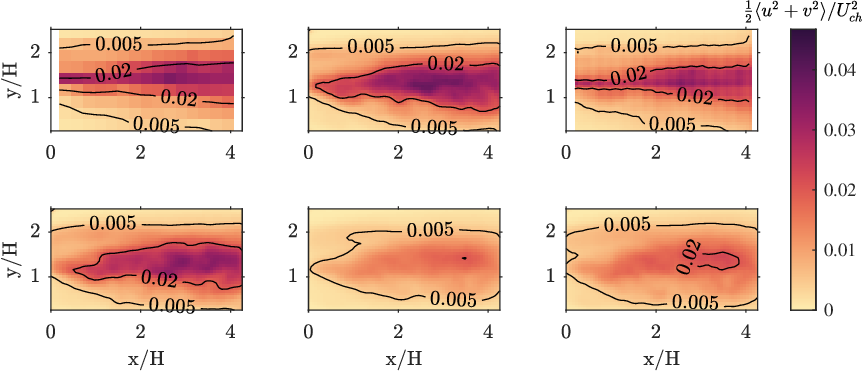}};

        % Approximate panel labels for Figure 2
        \node[anchor=north west] at ([xshift=0.7cm,yshift=0.15cm]img2.north west) {\textbf{a)} \revA{LR}};
        \node[anchor=north west] at ([xshift=5.1cm,yshift=0.15cm]img2.north west) {\textbf{b)} \revA{HR-LOR}};
        \node[anchor=north west] at ([xshift=9.5cm,yshift=0.15cm]img2.north west) {\textbf{c)} \revA{Cubic}};

        \node[anchor=north west] at ([xshift=0.7cm,yshift=-2.9cm]img2.north west) {\textbf{d)} \revA{KF}};
        \node[anchor=north west] at ([xshift=5.1cm,yshift=-2.9cm]img2.north west) {\textbf{e)} \revA{LSE}};
        \node[anchor=north west] at ([xshift=9.5cm,yshift=-2.9cm]img2.north west) {\textbf{f)} \revA{LSE+VR}};
    \end{tikzpicture}
    \caption{\revA{Spatial distribution of \ac{TKE} for the channel-flow case, normalized by the square of the free-stream channel flow velocity $U_{ch}$. Panels show: (a) \ac{LR} input; (b) \ac{HR}-\ac{LOR} reference; (c) cubic interpolation; (d) \ac{KF}; (e) \ac{LSE}; and (f) \ac{LSE}+\ac{VR}. Black isolines denote selected \ac{TKE} levels.}}
    \label{fig:tke_channel}
\end{figure*}
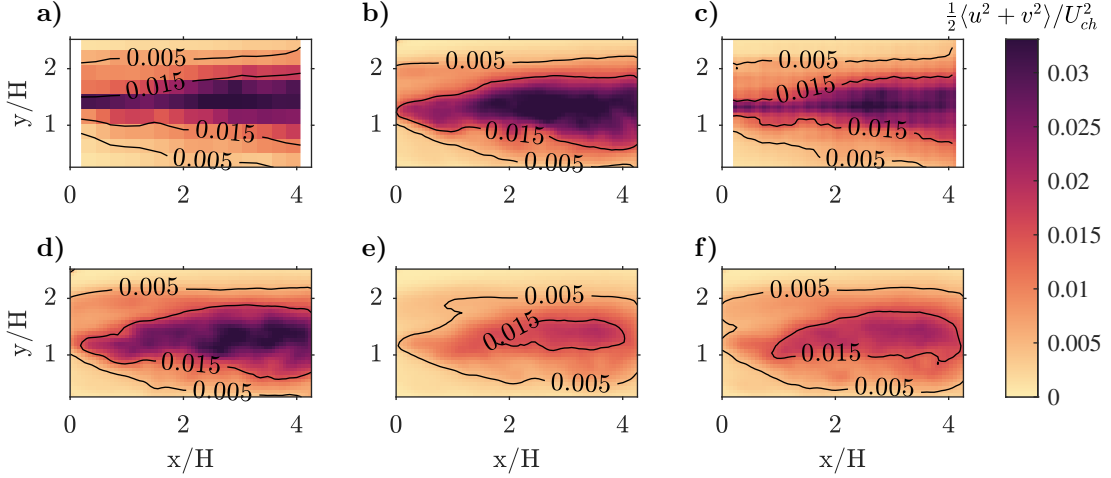

The spatial distribution of the in-plane \ac{TKE}, defined here as $\frac{1}{2}\langle u^2(\mathbf{x},t)+v^2(\mathbf{x},t)\rangle$ is reported in Figs.~\ref{fig:tke_jet} and \ref{fig:tke_channel} for the jet and channel-flow cases, respectively. %Compared with mean vorticity, \ac{TKE} is a more discriminating statistic, since it depends quadratically on the fluctuating velocity field and is therefore more sensitive to reconstruction errors affecting the fluctuation amplitude and spatial organization.

The \ac{HR}-\ac{LOR} reference reveals clear and physically consistent high-energy regions. In the jet, the \ac{TKE} is concentrated along the two shear layers, where the strongest fluctuations are associated with the roll-up and downstream convection of coherent structures. In the channel flow, the highest energy levels are found in the separated shear layer and in the downstream recovery region.

The three proposed estimators recover these energetic regions with good overall consistency. Among them, the \ac{KF} result is the closest to the reference in both spatial footprint and amplitude, with high-energy regions appearing at the correct locations and with a realistic spatial spreading. The \ac{LSE} reconstruction also captures the overall topology of the \ac{TKE} field, but the energetic regions appear smoother and less sharply defined, with reduced peak levels\revA{, indicating an attenuation of the turbulent fluctuations}. This behaviour is consistent with the tendency of linear regression estimators to smooth the fluctuating content, especially for less energetic and less correlated components of the flow. The \ac{LSE}+\ac{VR} correction mitigates this effect by restoring part of the missing modal variance, leading to a better recovery of the energy distribution while preserving a physically regular spatial organization.

The cubic-interpolation baseline exhibits a different trade-off. In terms of \ac{TKE} magnitude, it can provide a closer estimate than the LSE-based reconstructions in some regions, particularly in the channel-flow case, where the latter tend to underestimate the fluctuation energy. However, unlike the \ac{KF} estimate, which provides the closest agreement with the reference in both spatial distribution and amplitude, the cubic-interpolated fields exhibit periodic energy modulations that are absent in the reference. These appear as alternating bands or ripples of higher and lower \ac{TKE}, particularly along the jet shear layers and in the energetic region of the channel flow. The artifacts arise from grid imprinting: when under-resolved and noisy \ac{LR} fluctuating fields are interpolated onto a finer grid, the interpolation operator transfers the footprint of the original sampling pattern into the reconstructed field. Because \ac{TKE} is a quadratic quantity, these non-physical oscillations are amplified and become much more evident than in the instantaneous velocity snapshots.

Overall, these results highlight the different trade-offs between the reconstruction approaches. The \ac{KF} provides the best overall agreement with the reference \ac{TKE} field, recovering both its amplitude and spatial organization, while the \ac{LSE}-based estimators provide spatially regular reconstructions at the cost of some attenuation of the fluctuation energy. Cubic interpolation can preserve the overall \ac{TKE} level more closely in some regions, but introduces grid-imprinting artifacts that are absent from the reference.\\

\subsection{Temporal and spatial spectral content}
\label{sec:results_spectra}

The spectral fidelity of the reconstructed fields is assessed through pre-multiplied temporal and spatial spectra of the streamwise velocity fluctuation $u$. In the following, $\langle \cdot \rangle_T$ denotes temporal averaging and $\langle \cdot \rangle_{\Omega}$ denotes spatial averaging over the spatial subdomain $\Omega$.

\begin{figure*}
    \centering
    \includegraphics{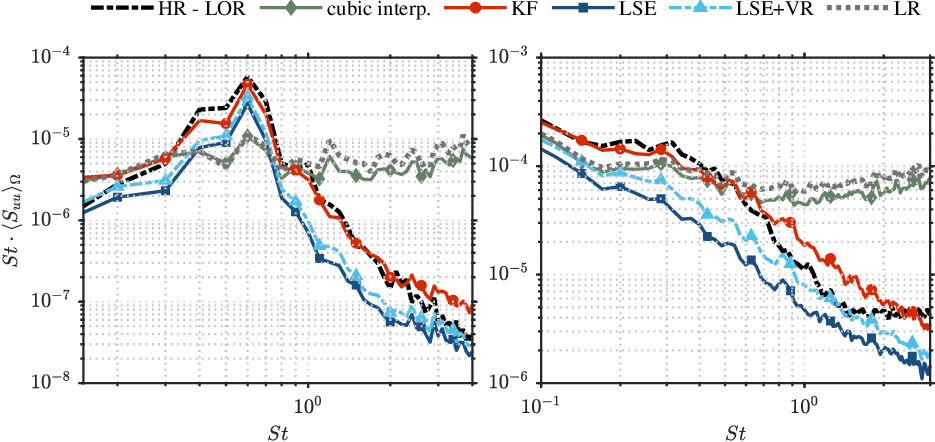}
    \caption{Premultiplied temporal \acp{PSD} of the streamwise velocity fluctuations for the jet (left) and channel-flow (right) cases, normalized respectively by $U_j$ and $U_{ch}$, reported as $St\,\langle S_{uu}\rangle_{\Omega}$. The spectra are spatially averaged over the subdomain $\Omega$ defined by $2 \leq x/D \leq 3$ and $-0.2 \leq y/D \leq 0.2$ for the jet, and $1 \leq x/H \leq 3.5$ and $1 \leq y/H \leq 2$ for the channel flow, and compared against the \ac{HR}-\ac{LOR} reference. \revA{The premultiplied spectra are reported in non-dimensional form, with the spectral energy normalized by $U_j^2$ and $U_{ch}^2$, respectively. The grey dotted line reports the spectra of the raw \ac{LR} fields.}}
    \label{fig:tpsd}
\end{figure*}

The quantity reported in the temporal spectral analysis of Fig.~\ref{fig:tpsd} is the premultiplied temporal \ac{PSD} of the streamwise velocity fluctuation, $St\,\cdot\langle  S_{uu}\rangle_{\Omega}$, computed from the local time signals within the subdomain $\Omega$, and the Strouhal number is defined as $St=fD/U_j$ for the jet and $St=fH/U_{ch}$ for the channel flow. The temporal spectra are then averaged over $\Omega$, defined as $2 \leq x/D \leq 3$ and $-0.2 \leq y/D \leq 0.2$ for the jet, and $1 \leq x/H \leq 3.5$ and $1 \leq y/H \leq 2$ for the channel flow, corresponding respectively to the jet core and the highly turbulent region behind the square rib. The corresponding subdomains are indicated by the orange dashed rectangles in Fig.~\ref{fig:exp_domains}.

\revA{The jet subdomain covers a comparatively compact region of size $1D \times 0.4D$, centred on the jet core, where the signature of the jet-column coherent dynamics emerges most clearly and the comparison between estimators is less affected by measurement noise.
The channel-flow subdomain is deliberately more extended: it samples a fully turbulent broadband region (representative of the behaviour that would also be observed across a turbulent shear layer) and the wider spatial averaging promotes the statistical convergence of the spectra, at the price of smoothing out strictly local contributions. It has been verified that the reported trends are insensitive to this choice: repeating the analysis over a much smaller nested subdomain leaves the relative behaviour of the estimators unchanged, at the cost of noisier spectra; the jet-case trends were likewise verified by recomputing the spectra within the shear-layer region. The two cases are therefore regarded as complementary: the former probes the recovery of coherent, spectrally localized dynamics, the latter the recovery of broadband turbulent content.}

\begin{figure*}
    \centering
    \includegraphics{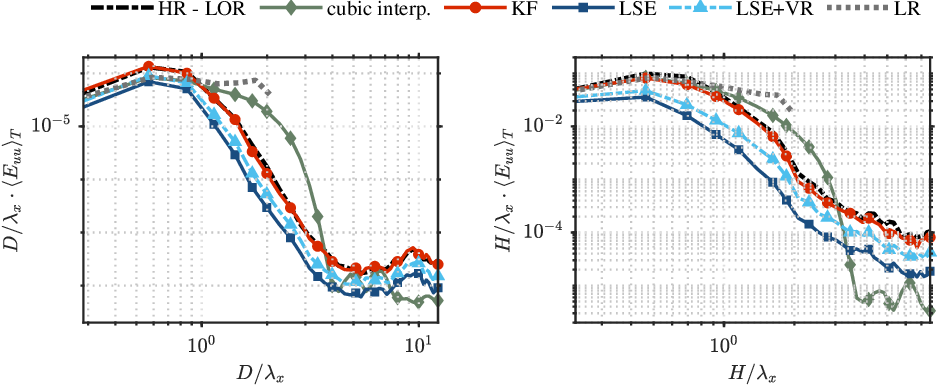}
   \caption{Premultiplied spatial spectra of the streamwise velocity fluctuations for the jet (left) and channel-flow (right) cases, normalized respectively by $U_j$ and $U_{ch}$, computed along the horizontal lines $y/D=0$ and $y/H=1.5$, respectively. The spectra are shown as $(D/\lambda_x)\,\langle E_{uu}\rangle_T$ for the jet and $(H/\lambda_x)\,\langle E_{uu}\rangle_T$ for the channel flow, and compared against the \ac{HR}-\ac{LOR} reference. \revA{The premultiplied spectra are reported in non-dimensional form, with the spectral energy normalized by $U_j^2$ and $U_{ch}^2$, respectively. The grey dotted line reports the spectra of the raw \ac{LR} fields, whose resolved wavenumber range terminates earlier owing to the coarser \ac{LR} vector spacing.}}
    \label{fig:spsd}
\end{figure*}

The premultiplied spatial spectra, shown in Fig.~\ref{fig:spsd}, are computed along a horizontal sampling line located at $y/D=0$ for the jet and at $y/H=1.5$ for the channel flow, as indicated by the grey horizontal line in Fig.~\ref{fig:exp_domains}. In this case, the spectra are represented in premultiplied non-dimensional form as $(L_{ref}/\lambda_x)\,\langle E_{uu}\rangle_T$, where $E_{uu}$ denotes the one-dimensional streamwise spatial spectrum of $u$, $L_{ref}$ is a reference length ($D$ for the jet and $H$ for the channel flow), and $\lambda_x$ is the spatial wavelength.

The temporal spectra show that all three proposed estimators \revA{capture the dominant frequency content of the flow}. In the jet case, the main spectral peak associated with the dominant coherent dynamics is recovered in both position and amplitude. In the channel-flow case, where the spectrum is broader and less sharply peaked, the low-frequency content is likewise captured correctly. The \ac{KF} estimator provides the closest agreement with the \ac{HR}-\ac{LOR} reference over the widest frequency range, including part of the high-frequency tail\revA{, capturing higher energy content with less attenuation, consistently with the \ac{TKE} results}. Standard \ac{LSE} and \ac{LSE}+\ac{VR} follow the same overall spectral shape and retain the dominant dynamics, but with a lower overall energy level, consistent with the attenuated fluctuating fields already observed in the \ac{TKE} maps. The \ac{LSE}+\ac{VR} is observed to perform \revA{consistently} better than the standard counterpart.

The cubic-interpolation baseline again exhibits the least reliable behaviour. Although it reproduces part of the low-frequency content \revA{(albeit with noticeable attenuation)}, it develops a markedly elevated broadband floor at intermediate and high Strouhal numbers. This excess floor is more than one order of magnitude above the reference over part of the spectrum, and in the jet tail it approaches nearly two orders of magnitude. As a result, the physical spectral decay is partially obscured and the identification of dominant peaks becomes less reliable, particularly in the channel-flow case where the broadband contamination masks part of the spectral structure. \revA{The spectra of the raw \ac{LR} fields, also reported in Fig.~\ref{fig:tpsd}, make the origin of this floor explicit: the \ac{LR} spectra follow the reference at low Strouhal numbers, but flatten into an elevated broadband floor at intermediate and high frequencies, which constitutes the spectral signature of the measurement noise of the lightweight single-pass processing. The cubic-interpolation baseline essentially inherits this floor, as expected from a purely spatial operator acting on the \ac{LR} fields, whereas the proposed estimators suppress it by constraining the reconstruction to the learned reduced-order subspace. These observations are consistent with the known spectral response of conventional \ac{PIV} \citep{foucaut2004}.}

The spatial spectra reinforce the same interpretation from a complementary perspective. The three data-driven estimators reproduce the \ac{HR}-\ac{LOR} spatial energy distribution over a broad range of resolved wavelengths, including part of the small-scale content that is absent from the raw \ac{LR} field. \revA{The raw \ac{LR} spectra, also included in Fig.~\ref{fig:spsd}, follow the reference over the largest resolved scales with a slight energy excess ascribable to measurement noise, but terminate at markedly smaller wavenumbers than all the other curves: the coarse \ac{LR} vector spacing bounds the shortest resolvable wavelength to the corresponding Nyquist limit, $\lambda_{x,\mathrm{min}} = 2\Delta x_{\mathrm{LR}}$, i.e., $D/\lambda_x \approx 2$ for the jet and $H/\lambda_x \approx 2$ for the channel flow. This early cut-off visualizes directly the spatial information gap that the proposed estimators are required to fill.} The \ac{KF} result provides the closest overall match and, especially in the jet case, nearly overlaps the \ac{HR}-\ac{LOR} reference over most of the resolved spectral range. The advantage of the \ac{VR} correction is particularly evident in this representation: compared with standard \ac{LSE}, \ac{LSE}+\ac{VR} \revA{recovers part of the missing energy} while preserving the correct spectral trend. In contrast, cubic interpolation departs from the reference already at intermediate wavelengths and \revB{exhibits characteristic undulations at wavelengths comparable to the original \ac{LR} vector spacing. In the jet case, this interpretation is quantitatively consistent with the \ac{LR} sampling pitch. From the image scaling, the \ac{LR} vector spacing is $\Delta x_{\mathrm{LR}} \approx 7.4~\mathrm{mm}$, corresponding to $D/\Delta x_{\mathrm{LR}} \approx 4.1$, namely about four \ac{LR} vectors across the nozzle diameter. 
The interpolated spectrum indeed departs from the reference around this scale, displaying an energy deficit (notch) near $D/\lambda_x \approx 4$ followed by a spurious energy excess at slightly smaller wavelengths, $D/\lambda_x \approx 5$--$6$.} These undulations result directly from grid imprinting and confirm that the interpolation baseline does not reconstruct the missing \ac{HR} content, but instead introduces \revB{artificial spatial scales} into the field.

From a physical standpoint, these spectra show that the proposed framework not only improves the visual quality of the reconstructed fields but also preserves the scale distribution of the fluctuating dynamics. This aspect is essential if the reconstructed fields are to be used for reduced-order modeling, system identification, or feedback-oriented state estimation.\\

\subsection{Reduced-order dynamics}
\label{sec:results_rom}
\begin{figure*}
    \centering
    \includegraphics{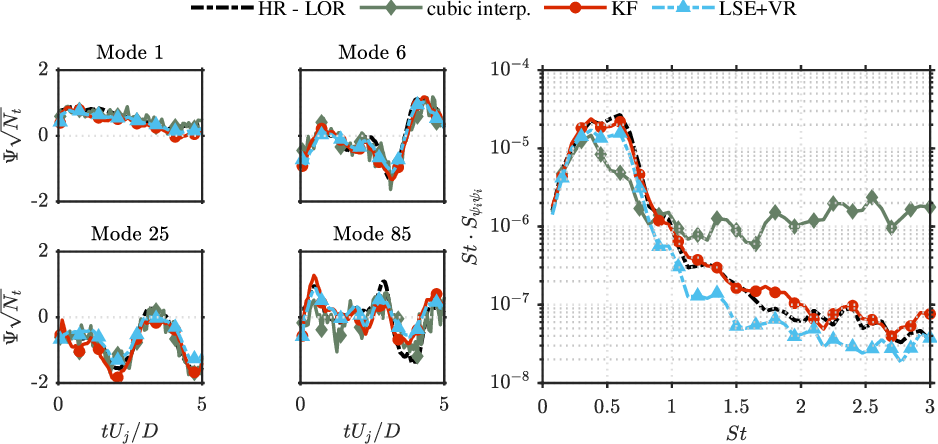}
    \caption{Jet-flow reduced-order dynamics. Left: time histories of selected \ac{POD} coefficients (modes 1, 6, 25, and 85) over five convective time unit. Right: premultiplied spectrum of the mode-6 coefficient. Coefficients are normalized as $\psi_i/\sqrt{N_t}$. For visual clarity, only the \ac{LSE}+\ac{VR} result is shown among the \ac{LSE}-based estimators.}
    \label{fig:psi_jet}
\end{figure*}

\begin{figure*}
    \centering
    \includegraphics{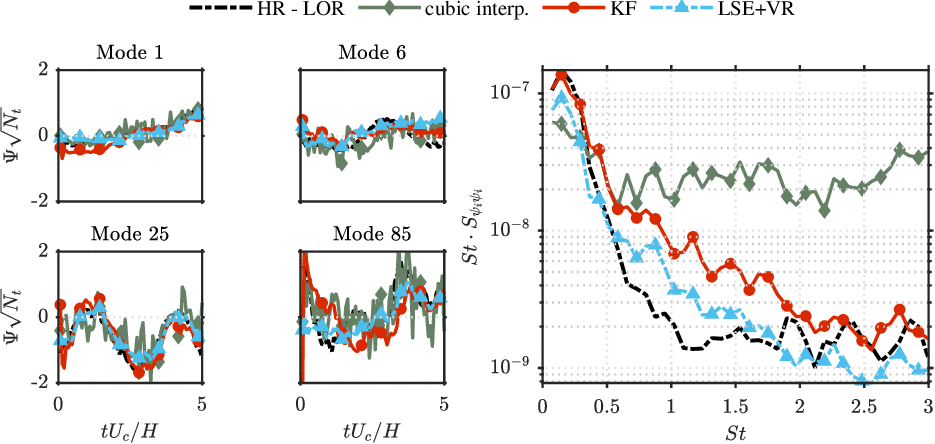}
    \caption{Channel-flow reduced-order dynamics. Left: time histories of selected \ac{POD} coefficients (modes 1, 6, 25, and 85) over five convective time unit. Right: premultiplied spectrum of the mode-6 coefficient. Coefficients are normalized as $\psi_i/\sqrt{N_t}$. For visual clarity, only the \ac{LSE}+\ac{VR} result is shown among the \ac{LSE}-based estimators.}
    \label{fig:psi_channel}
\end{figure*}

To assess the estimation accuracy directly in the reduced space, Figs.~\ref{fig:psi_jet} and \ref{fig:psi_channel} compare the time histories of four selected reduced-order coefficients together with the premultiplied spectrum of a representative one. The left-hand side of each figure reports the temporal evolution of the coefficients associated with modes $1$, $6$, $25$, and $85$ over a window of five convective time units. For each mode $i$, the plotted quantity is the corresponding scalar component of the reduced state, namely $\psi_{k,i}^{\mathrm{HR}}$ for the HR-LOR reference and $\hat{\psi}_{k|k,i}^{\mathrm{HR}}$ for the estimate, normalized by $\sqrt{N_t}$.

The convective time is defined as $t_c=tU_{ref}/L_{ref}$, with $U_{ref}=U_j$ and $L_{ref}=D$ for the jet, and $U_{ref}=U_{ch}$ and $L_{ref}=H$ for the channel flow. \revA{The four modes are selected to span the entire retained spectrum: the most energetic mode (mode 1), an intermediate-energy mode still dominated by coherent dynamics (mode 6), and two progressively higher-order, low-energy modes (modes 25 and 85), for which the estimation problem becomes increasingly challenging.} The right-hand side shows the premultiplied spectrum of the coefficient associated with mode~6\revA{, adopted as representative example because it combines a significant energy content with a well-defined spectral signature, which makes the differences between estimators clearly visible}. For visual clarity, only the \ac{LSE}+\ac{VR} result is shown among the \ac{LSE}-based estimators, as in this representation it largely overlaps with the standard \ac{LSE} result.

This analysis is particularly informative because the reduced coefficients are the actual state variables estimated by the proposed methods. Accurate coefficient reconstruction therefore implies not only a faithful reconstruction of the velocity field, but also a reliable low-dimensional representation of the underlying dynamics.

For the leading modes\revA{, i.e., the most energetic modes retrieved at the lowest mode indices}, all reconstruction methods recover the overall trend of the \ac{HR}-\ac{LOR} reference reasonably well. The \ac{KF} and \ac{LSE}+\ac{VR} estimates are generally the most consistent, reproducing the dominant low-order dynamics with good agreement in both amplitude and phase. Cubic interpolation often follows the large-scale trend as well, but already at low mode number, it exhibits visibly noisier coefficient traces, with superposed fluctuations that are absent from the reference. This behaviour is particularly evident in the channel-flow case, which is more strongly affected by measurement noise. As the mode rank increases, the reconstruction problem becomes progressively harder because the higher-order coefficients carry less energy and lie closer to the noise floor of the \ac{LR} measurements. In this regime, the interpolation-based estimates become much less trustworthy and may develop large spurious excursions, whereas the model-based estimators retain a meaningful correlation with the reference.

The attenuation of the \ac{LSE}-based reconstruction remains visible at higher mode number, as illustrated for instance by mode 85 in the channel-flow case. The \ac{VR} correction partially compensates for this loss by restoring part of the missing modal variance, yielding a reconstruction with amplitudes comparable with \ac{KF}. The coefficient spectra confirm the same interpretation. In the jet case, \ac{KF} reproduces the reference spectrum most closely over the widest Strouhal range. \revA{In the channel-flow case, the situation partially reverses: the \ac{KF} spectrum exceeds the reference at high frequencies, indicating that part of the measurement noise is transferred to the estimated coefficients through the correction step, whereas the smoother \ac{LSE}+\ac{VR} estimate remains in part closer to the reference over that range. This interchange in the relative performance of the two estimators between the two flow cases reflects the different role played by the linear dynamical model: in the jet, the quasi-periodic coherent dynamics are well captured by the one-step transition model, which allows the \ac{KF} to filter the measurement noise effectively; in the broadband channel flow, the linear model is less informative, and the \ac{KF} correction step consequently admits a larger share of the measurement noise. The same trend is observed across the other retained modes, with the \ac{KF} high-frequency excess visible in the channel-flow case over a wide range of mode indices, and the \ac{LSE}+\ac{VR} attenuation persisting at high mode index in both flow cases. It is worth remarking that both estimators consistently reproduce the \emph{shape} of the coefficient spectra (the spectral decay and the distribution of the dominant content) with the discrepancies essentially confined to the energy level, as in the \ac{LSE}+\ac{VR} attenuation. By contrast, the spectra of the cubic-interpolation baseline are nearly flat beyond the low-frequency range: the physical spectral decay is lost altogether, indicating that the interpolated coefficients are dominated by noise rather than by the underlying modal dynamics.} By contrast, cubic interpolation displays a clearly elevated broadband floor, confirming that it does not simply lose information, but also injects spurious modal activity.

\subsection{Temporal coherence from virtual probes}
\label{sec:results_autocorr}

\begin{figure*}
    \centering
    \includegraphics{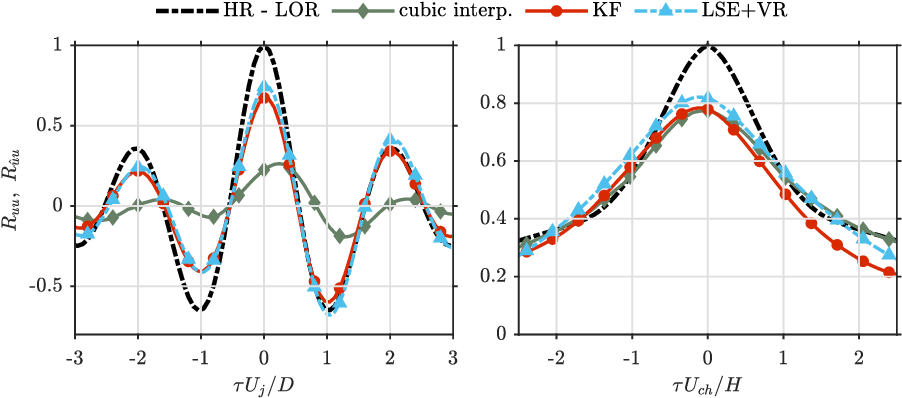}
    \caption{\revA{Lag-correlation analysis of the streamwise velocity fluctuation $u$ from virtual probes located at $(x/D,y/D)=(2,0)$ for the jet (left) and $(x/H,y/H)=(2,1.5)$ for the channel flow (right): temporal autocorrelation $R_{uu}$ for the \ac{HR}-\ac{LOR} reference, and normalized cross-correlation with the reference, $R_{\hat{u}u}$, for the estimators.} The time lag is normalized as $\tau U_j/D$ for the jet and $\tau U_{ch}/H$ for the channel flow. For visual clarity, only the \ac{LSE}+\ac{VR} result is shown among the \ac{LSE}-based estimators.}
    \label{fig:autocorr}
\end{figure*}

A complementary assessment of the reconstruction quality is provided by a lag-correlation analysis at a virtual probe located at $(x/D,y/D)=(2,0)$ for the jet and $(x/H,y/H)=(2,1.5)$ for the channel flow, respectively. \revA{For the \ac{HR}-\ac{LOR} reference, the temporal autocorrelation of the streamwise velocity fluctuation is reported,}
\begin{equation}
    R_{uu}(\tau)=\frac{\langle u(t)\,u(t+\tau)\rangle}{\langle u^2(t)\rangle},
\end{equation}
\revA{which is symmetric by construction. For the estimators, the normalized cross-correlation between the estimated and the reference probe signals is shown instead,
\begin{equation}
    R_{\hat{u}u}(\tau)=\frac{\langle \hat{u}(t)\,u(t+\tau)\rangle}{\sigma_{\hat{u}}\,\sigma_{u}},
\end{equation}
where $\sigma_{\hat{u}}$ and $\sigma_{u}$ denote the standard deviations of the estimated and reference signals. This cross-statistic is adopted deliberately, since it is the only one able to reveal a temporal misalignment between reconstruction and reference: an artificial time lag would displace the correlation peak away from $\tau=0$.} The results are reported in Fig.~\ref{fig:autocorr}. For visual clarity, only the \ac{LSE}+\ac{VR} result is shown among the \ac{LSE}-based estimators.

In the jet case, the \ac{HR}-\ac{LOR} reference displays the expected quasi-periodic \revA{symmetric} autocorrelation signature associated with the passage of coherent vortical structures and with a well-defined characteristic time scale. Both \ac{KF} and \ac{LSE}+\ac{VR} reproduce the delay structure well, without introducing an appreciable phase shift. \revA{The slight asymmetry of the reconstructed curves with respect to zero lag is an expected feature of this cross-statistic: the estimators are causal, operating on current and past measurements only, so their output does not correlate symmetrically with the past and the future of the reference signal. The correlation peak remains located at $\tau=0$, confirming that no appreciable artificial time lag is introduced.} \revA{In terms of correlation amplitude, \ac{LSE}+\ac{VR} performs comparably to \ac{KF}, following the reference more closely over part of the lag range.} By contrast, cubic interpolation yields a markedly damped and distorted autocorrelation shape, indicating that the interpolation procedure alters the underlying dynamics.

In the channel-flow case, the autocorrelation decays more smoothly and monotonically, consistent with the broader-band character of the separated-flow dynamics. The differences between the reconstruction methods are smaller than in the jet case. \ac{LSE}+\ac{VR} provides the best agreement in terms of correlation amplitude over much of the lag range, while cubic interpolation reproduces the overall \revA{correlation} shape reasonably well. In this representation, the \ac{KF} result does not exhibit a clear advantage over interpolation. While the Kalman filter enforces physical consistency, its reliance on a linear dynamical model leads to over-smoothing in the stochastic channel flow; here, local interpolation better captures the high-frequency fluctuations, despite lacking a physics-based formulation.

Overall, these results confirm that the data-driven estimators preserve the temporal organization of the reconstructed signal without introducing an evident artificial time lag, whereas cubic interpolation degrades the \revA{correlation} shape and therefore distorts the underlying temporal dynamics.

\subsection{Global reconstruction error}
\label{sec:results_error}

Finally, a compact scalar measure of reconstruction accuracy is reported in Table~\ref{tab:overall_error}. For the $n$-th snapshot and the $i$-th sampling point of the \ac{HR} grid, the local reconstruction error is defined as
\begin{equation}
e_{i,n}=\left\|\hat{\mathbf{u}}_n(\mathbf{x}_i)-\mathbf{u}^{\mathrm{LOR}}_n(\mathbf{x}_i)\right\|,
\end{equation}
where $\mathbf{x}_i$ denotes the location of the $i$-th \ac{HR} sampling point. \revA{Hatted quantities denote the estimated fields, while the superscript $\mathrm{LOR}$ refers to the \ac{HR}-\ac{LOR} reference used throughout the paper.} The overall error is then defined as the average normalized \ac{RMSE} of this local velocity error over the testing dataset:
\begin{equation}
\label{eq:delta}
\delta=\frac{1}{N_s\,U_{\mathrm{ref}}}\sum_{n=1}^{N_s}
\sqrt{\frac{1}{N_p^{\mathrm{HR}}}\sum_{i=1}^{N_p^{\mathrm{HR}}} e_{i,n}^2},
\end{equation}
where $N_s$ is the number of testing snapshots and $U_{\mathrm{ref}}$ denotes the reference velocity, namely $U_j$ for the jet and $U_{ch}$ for the channel-flow case.

The results reported in Table~\ref{tab:overall_error} show that\revA{, over the full resolved domain,} all three proposed estimators outperform the cubic-interpolation baseline in both configurations. Among the proposed methods, \ac{LSE} provides the lowest overall error in both the jet and channel-flow cases, with \ac{LSE}+\ac{VR} yielding nearly identical values. The improvement over cubic interpolation is appreciable, although still moderate in absolute terms. This should be interpreted in light of the intrinsic limitations of a pointwise scalar metric. By compressing all spatial and temporal information into a single number, $\delta$ cannot discriminate between physically consistent reconstructions and fields contaminated by non-physical artifacts. The metric $\delta$ should accordingly be regarded as a complementary summary indicator rather than the primary criterion for assessing reconstruction quality.

\revA{The apparent disparity between the field- and spectrum-based diagnostics, where \ac{KF} often provides the closest agreement with the reference, and the global error, where \ac{LSE} prevails, can be interpreted in relation to the nature of the estimators. The \ac{LSE} operator is identified by minimizing the mean-squared error over the training data, and is therefore near-optimal by construction with respect to $\delta$-type metrics; this optimality is achieved at the cost of a systematic attenuation of the fluctuations. The \ac{KF}, by contrast, trades pointwise accuracy for dynamic and spectral consistency, which is rewarded by the energy-based diagnostics but not by a pointwise scalar error. Additionally, to assess the sensitivity of $\delta$ to the choice of the spatial domain, the metric has also been evaluated restricting the spatial average to the most energetic regions of the flow, defined as the \ac{HR} grid points where the reference \ac{TKE} exceeds a fraction $\alpha$ of its maximum; the resulting values $\delta_{\alpha}$, together with the corresponding mask coverages, are reported in Table~\ref{tab:overall_error} for $\alpha=0.25$ and $\alpha=0.5$. The errors increase for all methods, as expected since the metric remains normalized by $U_{\mathrm{ref}}$ while the local fluctuation intensity grows, and the \ac{LSE}-based estimators remain the most accurate in the restricted domains as well, with \ac{LSE}+\ac{VR} marginally improving over standard \ac{LSE} in the channel-flow case. The contrast between the model-based estimators and the interpolation baseline, however, narrows in the energetic regions and, for the strictest threshold in the jet case ($\alpha=0.5$, isolating the thin high-\ac{TKE} bands of the shear layers over $8\%$ of the resolved points), the \ac{KF} pointwise error becomes slightly larger than that of interpolation. This behaviour reflects the different objective of the estimators discussed above: the \ac{KF} privileges dynamic and spectral consistency over pointwise optimality, retaining fluctuation amplitudes that a pointwise metric penalizes wherever energetic structures are slightly misplaced, whereas smoother estimates are favoured by $\delta$ even when they underrepresent the flow dynamics. The domain-restricted analysis thus provides quantitative confirmation of the complementary character of $\delta$-type indicators noted above.}

\revC{Finally, it is useful to quantify the error introduced by the low-order truncation alone to partially decouple the estimation error from the fidelity of the reference itself. Evaluated with the same metric of Eq.~\eqref{eq:delta}, but comparing the full \ac{HR} fields with their \ac{HR}-\ac{LOR} projection over the testing dataset, the truncation error amounts to $\delta_{\mathrm{LOR}} = 0.0550$ and $0.0540$  for the jet and channel-flow case, respectively. These values represent lower bounds for the total deviation of any of the proposed estimators from the full \ac{HR} fields and should be kept in mind when interpreting the errors of Table~\ref{tab:overall_error}, which are computed with respect to the \ac{HR}-\ac{LOR} reference. More precisely, since the estimated fields lie in the retained subspace by construction, while the truncation residual is orthogonal to it, the deviation of the proposed estimators from the full \ac{HR} fields is the root-sum-square of the two contributions, $\sqrt{\delta^2+\delta_{\mathrm{LOR}}^2}$, amounting for instance to $\approx 0.088$ and $\approx 0.117$ for \ac{LSE} in the jet and channel-flow cases, respectively; this decomposition does not apply to the interpolation baseline, whose output is not confined to the retained subspace. Since the truncation rank is selected by the \textit{elbow criterion} precisely so that the discarded modes are predominantly noise-dominated (Section~\ref{sec:formulation}), the \ac{HR}-\ac{LOR} reference acts as a denoised version of the \ac{HR} fields, and $\delta_{\mathrm{LOR}}$ partially reflects removed measurement noise; it should therefore be regarded as a conservative upper bound on the information excluded by the reference.}

% \begin{table}[t]
% \centering
% \caption{Normalized \ac{RMSE} $\delta$ (Eq.~\ref{eq:delta}) for the two 
%          flow cases. The best result in each column is highlighted in bold.}
% \label{tab:overall_error}
% \begin{tabular}{lcc}
% \toprule
% Method & Jet flow & Channel flow \\
% \midrule
% Cubic interpolation & 0.0876 & 0.1193 \\
% KF                  & 0.0792 & 0.1136 \\
% LSE                 & \textbf{0.0688} & \textbf{0.1042} \\
% LSE+VR              & 0.0695 & 0.1043 \\
% \bottomrule
% \end{tabular}
% \end{table}

\begin{table*}[t]
\centering
\caption{Normalized \ac{RMSE} $\delta$ (Eq.~\ref{eq:delta}) for the two flow cases\revA{, computed over the full resolved domain and restricted to the most energetic regions, $\delta_{\alpha}$, defined as the \ac{HR} grid points where the reference \ac{TKE} exceeds a fraction $\alpha$ of its maximum (the spatial average of Eq.~\ref{eq:delta} being restricted to those points). The last row reports the fraction of resolved points retained by each mask.} The best result in each column is highlighted in bold.}
\label{tab:overall_error}
\begin{tabular}{lcccccc}
\toprule
 & \multicolumn{3}{c}{Jet flow} & \multicolumn{3}{c}{Channel flow} \\
\cmidrule(lr){2-4}\cmidrule(lr){5-7}
Method & $\delta$ & \revA{$\delta_{0.25}$} & \revA{$\delta_{0.50}$} & $\delta$ & \revA{$\delta_{0.25}$} & \revA{$\delta_{0.50}$} \\
\midrule
Cubic interpolation & 0.0876 & \revA{0.1235} & \revA{0.1525} & 0.1193 & \revA{0.1501} & \revA{0.1672} \\
KF                  & 0.0792 & \revA{0.1230} & \revA{0.1632} & 0.1136 & \revA{0.1460} & \revA{0.1625} \\
LSE                 & \textbf{0.0688} & \revA{\textbf{0.1073}} & \revA{\textbf{0.1398}} & \textbf{0.1042} & \revA{0.1345} & \revA{0.1501} \\
LSE+VR              & 0.0695 & \revA{0.1075} & \revA{0.1407} & 0.1043 & \revA{\textbf{0.1338}} & \revA{\textbf{0.1492}} \\
\midrule
\revA{Mask coverage} & \revA{$100\%$} & \revA{$32\%$} & \revA{$8\%$} & \revA{$100\%$} & \revA{$51\%$} & \revA{$29\%$} \\
\bottomrule
\end{tabular}
\end{table*}

% \begin{figure}
%     \centering
%     \includegraphics{Fig/ErrorBarPlot_NoMeanFlow.eps}
%     \caption{Overall reconstruction error \(\delta\), defined as the normalized \ac{RMSE} of the fluctuation velocity magnitude over the testing dataset, for the jet and channel-flow cases.}
%     \label{fig:error_bar}
% \end{figure}

\subsection{Computational cost of the reduced-order refinement}
\label{sec:cost_rom_refinement}

To quantify the online overhead introduced by the proposed \ac{LR}-to-\ac{HR} refinement stage, a dedicated  microbenchmark implemented in \texttt{C++} was performed on the jet case. The benchmark was implemented following the online workflow of Method I (direct Kalman-filter estimation), including the learned state-transition matrix \(\mathbf{F}\). Although the exact latent-state update depends on the chosen estimator, the benchmark is intended to provide a representative order-of-magnitude assessment of the additional \revA{computational overhead of the reduced-order refinement}. The benchmark was carried out on a workstation running Microsoft Windows 11 Pro (build 26200), equipped with an AMD Ryzen Threadripper 7960X CPU (24 physical cores, 48 logical processors) and 256~GB of RAM. All reported timings correspond to single-threaded execution. The benchmark accounts only for the additional algebraic operations required once the \ac{LR} velocity field is already available, namely: projection of the incoming \ac{LR} field onto the reduced space, estimation of the \ac{HR} latent state, and reconstruction of the \ac{HR} field from the estimated latent coefficients.

\begin{figure}
    \centering
    \includegraphics{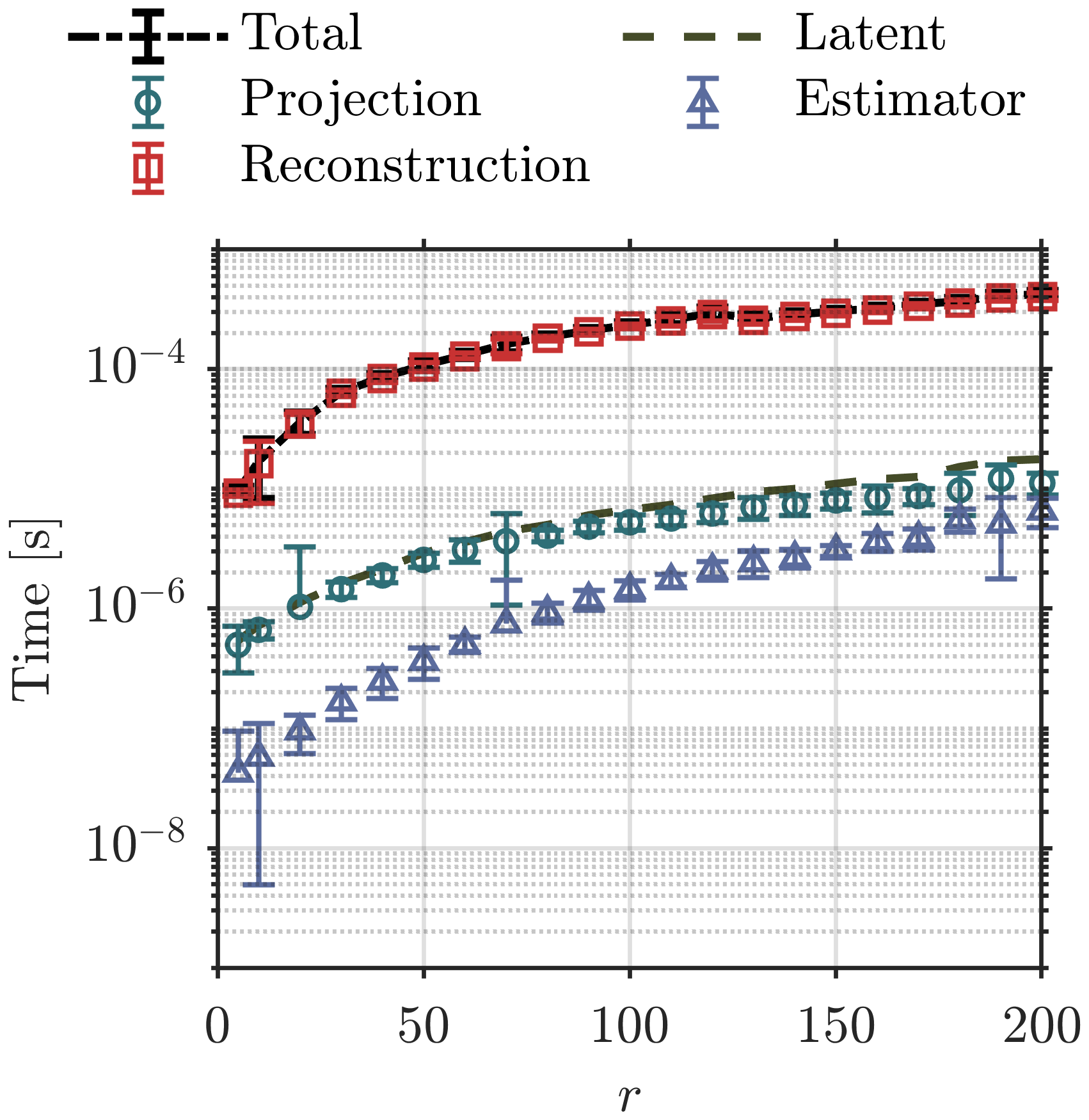}
    \caption{Computational cost of the additional online reduced-order refinement stage for the jet case as a function of retained rank \(r\). The reported timings distinguish the contribution of the \ac{LR} projection, latent-state estimation, and \ac{HR} field reconstruction, together with their total cost. The line labeled \emph{Latent} \revA{(i.e., operating only on the reduced-order \ac{POD} coordinates)} denotes the cost of operating only in the reduced space, i.e., projection plus latent-state estimation without full-field reconstruction. This quantity is particularly relevant for control-oriented implementations, in which the latent \ac{HR} representation is sufficient and the full reconstructed field is not required online. Error bars denote the standard deviation over repeated runs.}
    \label{fig:timing_rank}
\end{figure}

The results are reported in Fig.~\ref{fig:timing_rank}. For retained ranks up to \(r=200\), the total added cost remains below \(0.5\) ms on average, with worst observed values below \(0.8\) ms. In the jet case, this overhead is comfortably below the \(10\) ms time budget associated with the \(100\) Hz update rate, indicating that the reduced-order refinement is not the limiting factor for real-time deployment. In the channel-flow case, the available time budget would be more restrictive because of the higher acquisition frequency. On the other hand, the number of spatial degrees of freedom $N_p^{\mathrm{HR}}$ is lower by a few times, so the cost of the reduced-order operations is also expected to be correspondingly smaller.

The timing breakdown also clarifies the computational share of the steps of the process. The projection of the \ac{LR} field and the latent-state estimation remain comparatively inexpensive, whereas the dominant contribution is the reconstruction of the full \ac{HR} field. This trend is expected from the dimensionality of the three operations: the \ac{LR} projection scales as \(\mathcal{O}(r\,N_p^{\mathrm{LR}})\), the latent-state estimation as \(\mathcal{O}(r^2)\), and the full-field reconstruction as \(\mathcal{O}(N_p^{\mathrm{HR}}\,r)\). Since \(N_p^{\mathrm{HR}}\gg r\), the reconstruction step naturally dominates the total overhead.

This distinction is particularly relevant for control-oriented applications. When the objective is feedback estimation or low-order state monitoring, the online procedure can stop at the latent \ac{HR} representation, without reconstructing the full field at each time step. In Fig.~\ref{fig:timing_rank}, this case corresponds to the line labeled \emph{Latent}, which includes only the reduced-space operations and excludes full-field reconstruction. Its cost remains of the order of \(10\,\upmu\mathrm{s}\), well below the characteristic time required for the \ac{LR} velocity-field estimation itself. The proposed framework is therefore especially attractive when the reduced coordinates, rather than the full reconstructed field, are the quantities of primary interest.

\section{Conclusions}
\label{sec:Conclusions}

A data-driven reduced-order framework has been proposed to estimate, in real time, high-resolution velocity fields and reduced-order flow coordinates from coarse \ac{rtEBIV} measurements. The approach was assessed on two broadband turbulent-flow datasets, namely a submerged water jet and a channel flow over a square rib, using paired \ac{LR}/\ac{HR} data obtained from the same raw \ac{EBV} recordings through two processing pipelines of different fidelity. As a result, the offline training stage does not require any modification of the experimental setup or additional instrumentation, since the information used to learn the estimator is extracted from the same measurement chain that is employed online.

Compared with direct upsampling with cubic interpolation of the \ac{LR} measurements, the proposed estimators provide consistently more reliable reconstructions of fluctuation-related flow content. Their benefit is not limited to visually plausible super-resolution, but extends to a more physically consistent reconstruction of second-order statistics, temporal and spatial spectra, reduced-order modal dynamics, and temporal autocorrelation, while avoiding the grid-imprinting artifacts introduced by purely spatial upsampling. In this sense, the proposed framework should not be interpreted as a simple \ac{LR}-to-\ac{HR} interpolation operator, but rather as a reduced-order estimator that combines learned cross-resolution mappings with temporal regularization.

An important aspect of the methodology is that the reduced-order modeling stage also acts as a denoising mechanism. This is particularly relevant in the present context, where the online \ac{LR} measurements are obtained from a deliberately lightweight single-pass processing chain, which is computationally efficient but more sensitive to noise than the offline multi-pass procedure used to construct the \ac{HR} reference. \revA{The signature of this excess noise is directly visible in the spectral analyses of Section~\ref{sec:results_spectra}, where the raw \ac{LR} spectra exhibit an elevated broadband floor at high frequencies and wavenumbers (Figs.~\ref{fig:tpsd}--\ref{fig:spsd}).} Direct interpolation propagates this noisy and under-resolved information to the finer grid, whereas the proposed estimators constrain the reconstruction to a learned reduced-order subspace. This effect is especially valuable when the reduced-order \ac{HR} coordinates themselves are the quantities of interest, since operating directly in reduced space provides access to a compact and dynamically meaningful representation of the flow while filtering part of the measurement noise.

Across the different diagnostics, no single estimator dominates uniformly in every representation. \ac{LSE} yields the lowest global scalar error, \ac{LSE}+\ac{VR} remains very close to it and improves the recovery of fluctuation energy and higher-order content, while \ac{KF} remains robust in several field- and spectrum-based comparisons. The variance-rescaling correction should nevertheless be interpreted with care, since it is heuristic and identified from empirical statistics rather than derived from an optimal estimation principle. Its value is therefore mainly pragmatic: it compensates part of the systematic attenuation of standard \ac{LSE} without changing the overall reduced-order estimation framework.

From a practical standpoint, the additional online cost of the refinement stage remains small compared with the real-time \ac{LR} measurement itself. In particular, when only the reduced-order \ac{HR} representation is required, the method can operate entirely in a reduced space at very low cost. The presence of a learned dynamical model also provides an additional practical advantage: if an \ac{LR} measurement is temporarily unavailable, for example because of frame skipping, deliberate downsampling, or a transient processing bottleneck, the reduced \ac{HR} state can still be propagated for a limited time using the model prediction alone. Although such a prediction-only estimate is less reliable than a measurement-corrected one, it remains preferable to a complete loss of the state estimate over short horizons.

The present methodology nevertheless remains tied to the operating conditions represented in the training data. In its current form, it should therefore not be regarded as a general solution for actuation-modified or strongly off-design flow regimes. \revB{This limitation should, however, be put in perspective. First, the offline training stage is required only once: its role is analogous to that of a calibration procedure, after which the measurement chain delivers offline-grade reduced-order estimates at real-time rates that the offline pipeline could not sustain. The cost of this calibration is quantified in Appendix~\ref{app:trainsize}: training records of the order of $10^2$ convective time units suffice for the estimators to outperform the interpolation baseline, and about $300$ convective time units recover the full performance reported in Table~\ref{tab:overall_error}. This corresponds to less than one second of acquisition for the channel-flow case. The same train-offline/estimate-online paradigm underpins established real-time estimation frameworks such as SPPIV and its closed-loop applications \citep{kandaProofofconceptStudySparse2022,nonomura2025,viguera2026}. Second, the real-time monitoring and diagnostics of unactuated flows (e.g., continuous monitoring of experiments, detection of transients, or virtual sensing in reduced-order coordinates) constitutes a complete application domain in which the method is directly deployable in its present form. The one-off calibration is therefore performed only once before deployment: since tracking the reduced-order state requires storing only a few hundred coefficients per snapshot at negligible computational cost (Section~\ref{sec:cost_rom_refinement}), acquisitions substantially longer than the offline processing could afford become accessible, easing the convergence of statistics that require long records. Third, the framework is already applicable to a relevant class of flow-control problems, namely those in which the actuation does not significantly alter the dynamics within the sensed region. A notable example is the configuration in which sensing, actuation, and performance evaluation are located in distinct areas of the region of interest, so that the actuation affects the reward region but only marginally the sensed one, as in the real-time \ac{EBIV} control experiment presented by \citet{willert2026}.} Extending the framework toward actuation-aware estimation \revA{(i.e., toward fully closed-loop active-flow-control conditions in which the actuation modifies the sensed dynamics)} and multi-regime training is a natural next step\revB{; a promising route, common in the flow-control literature, is to train the estimator over a basis of spanned actuation policies, so that the Kalman filter operates on a reduced-order representation valid across the actuated regimes}. In addition, the present formulation relies on a linear reduced-order transition model and on a linear \ac{POD}-based encoding. Future developments may therefore explore nonlinear extensions of both ingredients, for example through operator-inference strategies and nonlinear filtering frameworks such as the extended or unscented Kalman filter\revA{, which accommodate nonlinear state-transition and measurement models through local linearization and deterministic sampling of the state statistics, respectively}, as well as nonlinear reduced-order representations based on autoencoders or related machine-learning approaches. \revA{Finally, the development of a validated synthetic continuous- and pulsed-\ac{EBIV} simulator (capable of reproducing the sensor behaviour under pulsed illumination of tracer particles, including per-pixel threshold dispersion, timing jitter, and arbiter saturation) would provide access to ground-truth-based assessments and represents, in the authors' view, a valuable development for the event-based velocimetry community at large.}

Within its current scope, however, the results demonstrate that deliberately coarse \ac{rtEBIV} processing can be used to extend the effective real-time operating range of the measurement chain toward higher sampling frequencies, while still recovering substantially richer and dynamically more consistent \ac{HR} flow representations through reduced-order estimation. This is promising not only for high-quality real-time flow diagnostics, but also as a first step toward future observer-based and closed-loop flow control applications.

% \section*{Author contributions}

% L.F. and S.D. conceptualized the study. S.D. was responsible for funding acquisition and project administration. S.D., M.R., and G.C. supervised the work, while S.D. and C.E.W. provided the necessary resources. L.F. curated the data. L.F., E.A., and C.E.W. conducted the experiments and processed the experimental data. L.F., E.A., and S.D. performed the formal analysis. L.F., E.A., M.R., and S.D. developed the methodology. C.E.W. developed the EBIV processing software, while L.F. and E.A. implemented the high-resolution estimation framework. L.F. wrote the original draft of the manuscript. All authors reviewed and edited the manuscript.

\section*{CRediT author statement}

\textbf{L.F.:} Conceptualization, Data curation, Formal analysis, Investigation, Methodology, Software, Writing--original draft, Writing--review and editing. 
\textbf{E.A.:} Formal analysis, Investigation, Methodology, Software, Writing--review and editing. 
\textbf{C.E.W.:} Investigation, Resources, Software, Writing--review and editing. 
\textbf{M.R.:} Methodology, Supervision, Writing--review and editing. 
\textbf{G.C.:} Supervision, Writing--review and editing. 
\textbf{S.D.:} Conceptualization, Formal analysis, Funding acquisition, Methodology, Project administration, Resources, Supervision, Writing--review and editing.

\section*{Declaration of competing interest}
The authors declare that they have no known competing financial interests or personal relationships that could have appeared to influence the work reported in this paper.

\section*{Code and data availability}

Data supporting this work are available on Zenodo at the following DOI: \href{10.5281/zenodo.20037404}{10.5281/zenodo.20037404}. The archive includes the \texttt{.raw} event-camera recordings, a subset of the corresponding pseudo-snapshots, and the processed velocity fields used in the present analysis. Additional data are available from the corresponding author upon reasonable request.

The real-time \ac{EBIV} workflow and the high-resolution estimation framework proposed in this work are implemented in \emph{VIBEstream}, an open-source Python package available at \href{https://github.com/LFranceschelli/VIBEstream}{https://github.com/LFranceschelli/VIBEstream}. VIBEstream provides a simple interface to operate an event camera connected to the host computer: recording and storage of raw event streams and pseudo-images, live streaming and monitoring of the scene, quick \ac{PIV}-style processing, and real-time \ac{EBIV} with direct output of the velocity-field information, so that it can be readily integrated in laboratory experiments and control loops. The package is intended as an educational, easily extensible implementation, and also includes the high-resolution estimation methods presented in this paper; a faster and more robust C++ version, used for the dedicated real-time benchmark of Section~\ref{sec:cost_rom_refinement}, is under development and is shared under collaboration agreements.

\section*{Acknowledgements}
This project has received funding from the European Research Council (ERC) under the European Union’s Horizon 2020 research and innovation programme (grant agreement No 949085, NEXTFLOW ERC StG). Views and opinions expressed are however those of the authors only and do not necessarily reflect those of the European Union or the European Research Council. Neither the European Union nor the granting authority can be held responsible for them.

The help of Michael Schroll of the DLR Institute of Technology was invaluable in setting up the air flow experiment.

\section*{Declaration of generative AI and AI-assisted technologies in the writing process}
During the preparation of this work, the authors used Chat-GPT and Grammarly to improve the readability and language of this manuscript. After using this tool/service, the authors reviewed and edited the content as needed and took full responsibility for the publication’s content.

\begin{appendices}

\section{Sensitivity to the training-dataset size}
\label{app:trainsize}

\revB{The offline training stage of the proposed framework plays the role of a calibration procedure, and its practical cost is set by the length of the recording required to identify the reduced-order bases and operators. This appendix quantifies this cost by repeating the complete training procedure on progressively shorter portions of the training data.}

\revB{For each training-set size $N_{train} \in \{500, 1000, 1500, 2250, 3000, 4500\}$, the \ac{KF} and \ac{LSE} estimators are trained on contiguous windows of snapshots. Five window positions are considered within the full training range, except for the largest training-set size, for which only one window is available. For each window, the estimators are built using only the information available within that window, namely the mean flow, the \ac{POD} bases, the truncation ranks selected with the \textit{elbow criterion} described in Section~\ref{sec:formulation}, the mapping operators, and the noise covariances. The validation and testing subsets are kept fixed. Performance is evaluated using the metric of Eq.~\eqref{eq:delta} against the same \ac{HR}-\ac{LOR} reference used throughout the paper, which is built from the full training set. Therefore, $\delta(N_{train})$ quantifies how closely a shorter calibration reproduces the full-training product. The comparison is performed on the total velocity fields, so that errors in the mean-flow estimation from shorter records are also included.}

\begin{figure*}
    \centering
    \includegraphics{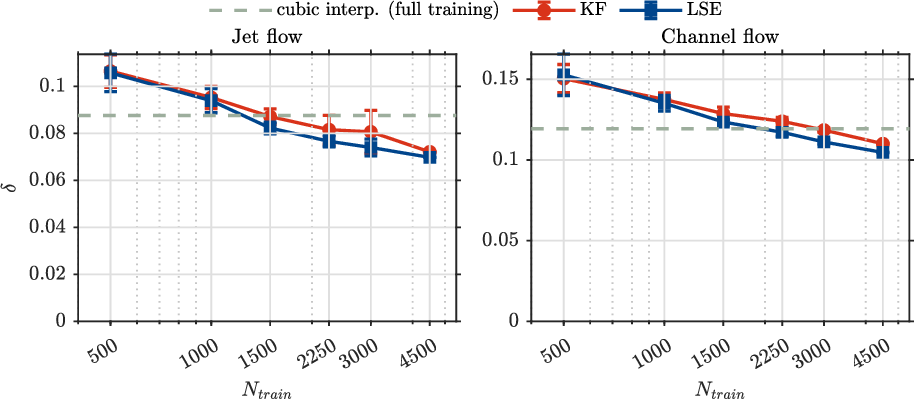}
    \caption{\revB{Global reconstruction error $\delta$ (Eq.~\ref{eq:delta}, evaluated against the full-training \ac{HR}-\ac{LOR} reference) as a function of the training-dataset size, for the jet (left) and channel-flow (right) cases. Symbols and error bars denote the mean and standard deviation over the five training-window positions. The dashed line reports the full-training cubic-interpolation baseline of Table~\ref{tab:overall_error}.}}
    \label{fig:trainsweep}
\end{figure*}

\revB{The results are reported in Fig.~\ref{fig:trainsweep}. The error decreases monotonically with the training-set size, with a limited scatter across the window positions, and the two estimators behave very similarly. Both match the full-training cubic-interpolation baseline with training records of about $1500$ snapshots for the jet and $2250$--$3000$ for the channel flow: expressed in physical terms, records of the order of $10^2$ convective time units in both cases (i.e., $15$--$30\,\mathrm{s}$ of acquisition for the jet and $0.3$--$0.6\,\mathrm{s}$ for the channel flow). Approximately $300$ convective time units recover the full-training performance of Table~\ref{tab:overall_error}, corresponding to $45\,\mathrm{s}$ of acquisition for the jet and to less than one second ($0.9\,\mathrm{s}$) for the channel flow. At the shortest records considered ($\approx 33$ convective time units), the estimation degrades gracefully, with $\delta$ exceeding the interpolation baseline by roughly $20$--$30\%$.}

\revB{Two remarks are due for a fair reading of these curves. First, $\delta$ is the most severe possible metric for the short-training estimators, since it combines the pointwise character of the metric discussed in Section~\ref{sec:results_error} --- which favours smoother estimates over dynamically consistent ones --- with the representational mismatch between the coarser training-window basis and the fixed full-training reference basis; the values at small $N_{train}$ should therefore be regarded as conservative bounds on the practical usability. Second, the truncation rank selected by the \textit{elbow criterion} can oscillate significantly between training windows, with only a marginal impact on $\delta$, indicating that the performance of the framework is robust with respect to the exact truncation choice. Overall, these results support the interpretation of the offline training stage as a lightweight, one-off calibration: a recording of the order of $10^2$ convective time units suffices to deploy estimators that outperform direct interpolation, with performance saturating toward the full-training level as the calibration record grows.}

\end{appendices}

\bibliography{Paper_HR-EBIV}

\end{document}